\documentclass[10pt]{artikel3}
\usepackage{amsfonts,amsmath,amssymb}
\usepackage{float,braket}
\usepackage{pslatex}
\usepackage{subfigure}
\usepackage{graphicx,wrapfig}
\usepackage[parfill]{parskip}
\usepackage[small,bf]{caption}
\newcommand{\gf}{\ensuremath{\mathrm{gf}}}
\newcommand{\YM}{\ensuremath{\mathrm{YM}}}
\newcommand{\e}{\ensuremath{\mathrm{e}}}
\newcommand{\s}{\ensuremath{\mathrm{s}}}
\newcommand{\p}{\partial}
\newcommand{\h}{\ensuremath{\mathrm{h}}}
\newcommand{\RGZ}{\ensuremath{\mathrm{RGZ}}}
\newcommand{\GZ}{\ensuremath{\mathrm{GZ}}}
\newcommand{\R}{\ensuremath{\mathrm{R}}}
\newcommand{\phys}{\ensuremath{\mathrm{phys}}}
\newcommand{\glue}{\ensuremath{\mathrm{glue}}}
\newcommand{\glueYM}{\ensuremath{\mathrm{YMglue}}}
\newcommand{\Rglue}{\ensuremath{\mathrm{Rglue}}}
\newcommand{\en}{\ensuremath{\mathrm{en}}}

\newcommand{\ext}{\ensuremath{\mathrm{ext}}}
\renewcommand{\d}{\ensuremath{\mathrm{d}}}

\begin{document}
\title{{\bf A renormalization group invariant scalar glueball operator in the (Refined) Gribov-Zwanziger framework.}}
\author{D.~Dudal \thanks{david.dudal@ugent.be}\,\,$^a$, S.~P.~Sorella\thanks{sorella@uerj.br}\,\,$^b$, N.~Vandersickel \thanks{nele.vandersickel@ugent.be}\,\,$^a$, H.~Verschelde\thanks{henri.verschelde@ugent.be}\,\,$^a$\\
\\
\small $^a$ \textnormal{Ghent University, Department of Mathematical Physics and Astronomy} \\
\small \textnormal{Krijgslaan 281-S9, 9000 Gent,Belgium}\\
\\
\small $^b$  \textnormal{Departamento de F\'{\i }sica Te\'{o}rica, Instituto de F\'{\i }sica, UERJ - Universidade do Estado do Rio de Janeiro}\\
\small   \textnormal{Rua S\~{a}o Francisco Xavier 524, 20550-013 Maracan\~{a}, Rio de Janeiro, Brasil }\normalsize}

\date{}
\maketitle

\begin{abstract}
This paper presents a complete algebraic analysis of the renormalizability of the $d=4$ operator $F^2_{\mu\nu}$ in the Gribov-Zwanziger (GZ) formalism as well as in the Refined Gribov-Zwanziger (RGZ) version. The GZ formalism offers a way to deal with gauge copies in the Landau gauge. We explicitly show that $F^2_{\mu\nu}$ mixes with other $d=4$ gauge variant operators, and we determine the mixing matrix $Z$ to all orders, thereby only using algebraic arguments. The mixing matrix allows us to uncover a renormalization group invariant including the operator $F^2_{\mu\nu}$. With this renormalization group invariant, we have paved the way for the study of the lightest scalar glueball in the GZ formalism. We discuss how the soft breaking of the BRST symmetry of the GZ action can influence the glueball correlation function. We expect non-trivial mass scales, inherent to the GZ approach, to enter the pole structure of this correlation function.
\end{abstract}

\section{Introduction}
QCD is the theory of strong interactions describing quarks and gluons which displays confinement at low energies. The mechanism behind confinement is still not successfully described. Even if one omits the quarks, the theory remains confining. Therefore, confinement is highly entangled with the dynamics of gluons, which makes glueballs very interesting objects to investigate. The existence of glueballs would be a pinnacle of the correctness of QCD, however, so far, there is still no clear experimental evidence for the existence of glueballs. This is mainly due to the mixing of glueball states with meson states which contain quarks. By increasing the statistics and/or by doing more involved experiments creating certain glueball states which cannot mix with quark states (oddballs), one hopes to uncover some clear evidence for glueball states. We mention a few experiments to demonstrate the general interest in glueballs: $\overline{\mathrm{P}}$ANDA \cite{Bettoni:2005ut}, BES III \cite{Chanowitz:2006wf} and GlueX \cite{Carman:2005ps}, ALICE at CERN \cite{Alessandro:2006yt}.  \\
\\
The lack of experimental evidence has not stopped the community to widely investigate glueballs in various theoretical models, see \cite{Mathieu:2008me} and their references therein. Currently, theoretical estimates of e.g.~masses of the different glueballs are compared to the lattice data. In lattice gauge theories, there is no doubt about the existence of glueballs and one can even work in pure Yang-Mills gauge theory \cite{Teper:1998kw}. There are many phenomenological models which contribute to our intuition in glueballs. More direct contact with fundamental QCD can be made by identifying suitable gauge invariant operators, which carry the correct quantum numbers to create/annihilate particular glueball states \cite{West:1996du}. This is in accordance with the direct approach to study bound states in quantum field theory \cite{Zimmermann:1958hg}. The mass of the glueball can then be determined by the leading singularity in its propagator which, if the glueball is stable, is just a simple pole. Of course, it is necessary to take into account non-perturbative effects, as glueballs are inherently connected to the non-perturbative region of QCD. One widely used method to estimate these propagators is based on QCD sumrules \cite{Novikov:1979va,Narison:2008nj}, while taking into account condensates, sometimes in combination with instanton or other nonperturbative effects. Also in holographic descriptions of QCD, such glueball correlators have already been investigated, see for instance \cite{Brower:2000rp,Forkel:2007ru}. \\
\\
In this paper, we shall concentrate on identifying a suitable composite operator $\mathcal R$, which is a renormalization group invariant containing $F^2_{\mu\nu}$, representing the lightest scalar glueball. Let us explain how we shall take into account a particular source of non-perturbative effects. For this, we need a bit of background. As is well known, the Faddeev-Popov quantization of the Yang-Mills gauge theory was constructed in order to restrict the path integration only over gauge inequivalent fields. This restriction is translated at the level of the action by implementing a gauge, e.g.~the Landau gauge $\p_\mu A_\mu = 0$, through the introduction of extra terms in the action, which in return break the local gauge invariance. In 1977, Gribov showed \cite{Gribov:1977wm} that this gauge fixing procedure in Yang-Mills gauge theories does not entirely restrict the path integration to gauge inequivalent fields, i.e.~there are still multiple gauge copies $A_\mu$ which all fulfill the Landau gauge condition. Moreover, it appeared that the infrared behavior of the gluon and the ghost propagator is strongly influenced when handling these copies. Therefore, there was a need for a formalism which took into account these Gribov copies, even if it would be only in a partial way.  After a semiclassical treatment by Gribov in \cite{Gribov:1977wm}, Zwanziger managed to construct an action which analytically implements the restriction to the Gribov region $\Omega$ \cite{Zwanziger:1989mf}. This action is called the Gribov-Zwanziger action $S_\GZ$. The region $\Omega$ is defined as the set of field configurations fulfilling the Landau gauge condition and for which the Faddeev-Popov operator,
\begin{eqnarray}
\mathcal{M}^{ab} &=&  -\p_\mu \left( \p_{\mu} \delta^{ab} + g f^{acb} A^c_{\mu} \right) \,,
\end{eqnarray}
is strictly positive. Therefore,
\begin{eqnarray}
\Omega &\equiv &\{ A^a_{\mu}, \, \p_{\mu} A^a_{\mu}=0, \, \mathcal{M}^{ab}  >0  \} \,.
\end{eqnarray}
The boundary, $\partial \Omega$, of the region $\Omega$ is called the (first) Gribov horizon. The restriction of the path integral to $\Omega$ removes most of the Gribov copies in the Landau gauge related to (infinitesimal) gauge transformations \cite{Gribov:1977wm}. However, there are still copies present in $\Omega$ and hence a further restriction to the Fundamental Modular Region (FMR), the region free of any Gribov copies, should be implemented. Unfortunately, till now, nobody knows how to handle such a restriction to the FMR. Therefore, the best analytical approach to restrict the number of gauge copies is by working with $S_\GZ$. We recall that $S_\GZ$ is renormalizable to all orders \cite{Zwanziger:1992qr,Maggiore:1993wq,Dudal:2005na}, even in the presence of massless \cite{Gracey:2005cx,Gracey:2006dr} or massive quarks \cite{Ford:2009ar}. Implementing the restriction to the horizon introduces a first non-perturbative mass scale, the so-called Gribov parameter $\gamma^2$. Also, we have found in \cite{Dudal:2007cw,Dudal:2008sp} that the auxiliary fields introduced by Zwanziger to construct the action $S_\GZ$, develop their own dynamics. This can introduce a second mass scale into the action. Generally, such non-perturbative mass scales are expected to be transmitted into the pole mass of the correlation functions.\\
\\
In a previous paper \cite{Dudal:2008tg} we have investigated the operator $F^2_{\mu\nu}$ in the ordinary Yang-Mills theory with Landau gauge fixing. This was already far from being trivial as at the quantum level mixing occurs with two other 4 dimensional operators, i.e.~a BRST exact operator $\mathcal E = s(\ldots)$, and an operator $\mathcal H$ which vanishes upon using the equations of motion. We have shown that this mixing does not have consequences when turning to physical states. Indeed, a BRST exact operator is always irrelevant at the level of physical states as the Yang-Mills action is invariant under the BRST symmetry. In this paper, we shall elaborate on the operator $F^2_{\mu\nu}$ by investigating it in the more complex Gribov-Zwanziger framework, whereby exploiting the construction we have set up in \cite{Dudal:2008tg}. In this case, a similar mixing shall occur, but, in contrast with the Yang-Mills case this mixing shall have consequences at the physical level. Indeed, as the Gribov-Zwanziger action gives rise to a soft breaking of the BRST symmetry \cite{Dudal:2008sp}, one can figure out that the corresponding BRST exact operator which will mix with $\mathcal F^2_{\mu\nu}$, will no longer be irrelevant. Let us mention that an attempt to calculate the glueball correlator $\Braket{F^2_{\mu\nu}(x) F^2_{\alpha\beta}(y)}$ has been done in \cite{preprint}, but without taking into account the mixing of $F^2_{\mu\nu}$ with other operators. We start the paper with an overview of the Gribov-Zwanziger action in section 2. We also recapitulate the Refined Gribov-Zwanziger action which takes into account the dynamics of the new fields introduced by Zwanziger. In section 3, a renormalizable action including the local, non-integrated operator $F^2_{\mu\nu}(x)$ is constructed whereby in section 4 we shall analyze the mixing of this operator to all orders. In section 5, we shall determine the renormalization group invariant which contains $F^2_{\mu\nu}$. We end this paper with a conclusion in section 6, where we also present some insights on the potential relevance of the soft BRST symmetry breaking of the GZ action.\\

\section{Overview of the (Refined) Gribov-Zwanziger action }
\subsection{The original Gribov-Zwanziger action}
In this section we shall shortly recapitulate the ordinary Gribov-Zwanziger action in Euclidean space time which implements the restriction of the path integral to the region $\Omega$. In \cite{Zwanziger:1989mf}, Zwanziger derived the following action,
\begin{eqnarray}
S_\h&=& S_{\YM} + S_{\gf} + \gamma^4 \int \d^d x\, h(x) \,,
\end{eqnarray}
with $S_{\YM}$ the classical Yang-Mills action,
\begin{eqnarray}
S_{\YM} &=& \frac{1}{4}\int \d^d x F^a_{\mu\nu} F^a_{\mu\nu} \,,
\end{eqnarray}
$S_{\gf}$ the Faddeev-Popov gauge fixing
\begin{eqnarray}
 S_{\gf} &=& \int \d^d x\,\left( b^a \p_\mu A_\mu^a +\overline c^a \p_\mu D_\mu^{ab} c^b \right)\,,
\end{eqnarray}
and $h(x)$ the horizon function,
\begin{eqnarray}
 h(x) &=& g^2 f^{abc} A^b_{\mu} \left(\mathcal{M}^{-1}\right)^{ad} f^{dec} A^e_{\mu} \,.
\end{eqnarray}
The horizon condition:
\begin{eqnarray}\label{horizonconditon}
\braket{h(x)} &=& d (N^2 -1) \,,
\end{eqnarray}
with $d$ the number of space-time dimensions, needs to be fulfilled in order to assure that we are working with a gauge theory quantized in the Landau gauge. This was proven using statistical arguments in \cite{Zwanziger:1989mf,Zwanziger:1992qr}. The action $S_\h$ contains a non-local term, but one can localize the horizon function by introducing the following set of additional fields: $\left(\overline \varphi_\mu^{ac}, \varphi_\mu^{ac}\right)$ which is a pair of complex conjugate bosonic fields, and $\left( \overline \omega_\mu^{ac},\omega_\mu^{ac} \right)$, which is a pair of Grasmann fields. After this procedure, $S_\h$ gets replaced by $S_\GZ$, which reads
\begin{eqnarray}\label{SGZphys}
S_\GZ &=& S_{0} +  S_{\gamma} \,,
\end{eqnarray}
with
\begin{eqnarray}\label{SGZ}
S_0 &=& S_\YM + S_\gf  \nonumber \\
&& + \int \d^d x\left( \overline \varphi_\mu^{ac} \p_\nu \left(\p_\nu \varphi_\mu^{ac} + g f^{abm} A_\nu^b \varphi_\mu^{mc} \right) - \overline \omega_\mu^{ac} \p_\nu \left( \p_\nu \omega_\mu^{ac} +g f^{abm} A_\nu^b \omega_\mu^{mc} \right) -g \left( \p_\nu \overline \omega_\mu^{ac} \right) f^{abm} \left( D_\nu c \right)^b \varphi _{\mu}^{mc}\right)  \,, \nonumber\\
S_{\gamma}&=& -\gamma ^{2} g \int \d^d x\left( f^{abc}A_\mu^a \varphi_\mu^{bc} + f^{abc} A_\mu^a \overline \varphi_\mu^{bc} + \frac{d}{g} \left(N^{2}-1\right)  \gamma^2 \right) \,,
\end{eqnarray}
We can further simplify the notation of the additional fields $\left( \overline \varphi_\mu^{ac},\varphi_\mu^{ac},\overline \omega_\mu^{ac},\omega_\mu^{ac}\right) $ as $S_0$ displays a symmetry with respect to the composite index $i=\left( \mu,c\right)$. Therefore, we can set
\begin{equation}
\left( \overline \varphi_\mu^{ac},\varphi_\mu^{ac},\overline \omega_\mu^{ac},\omega_\mu^{ac}\right) =\left( \overline \varphi_i^a,\varphi_i^a,\overline \omega_i^a,\omega_i^a \right)\,,
\end{equation}
so we get
\begin{eqnarray}
S_{0}&=&S_\YM + S_\gf + \int \d^d x \left( \overline \varphi_i^a \p_\mu \left( D_\mu^{ab} \varphi^b_i \right) - \overline \omega_i^a \p_\mu \left( D_\mu^{ab} \omega_i^b \right) - g f^{abc} \p_\mu \overline \omega_i^a    D_\mu^{bd} c^d  \varphi_i^c \right) \,.
\end{eqnarray}
Finally, the horizon condition \eqref{horizonconditon} can be written in a more practical version as
\begin{eqnarray}
\frac{\p \Gamma}{\p \gamma^2} &=& 0\,,
\end{eqnarray}
whereby the quantum action $\Gamma$ is obtained through the definition
\begin{eqnarray}
\e^{-\Gamma} &=& \int [\d\Phi] \e^{-S_\GZ} \,,
\end{eqnarray}
where $\int [\d\Phi]$ stands for the integration over all the fields.\\
\\
For the Gribov-Zwanziger action, the conventional BRST symmetry is softly broken \cite{Zwanziger:1989mf,Dudal:2008sp}. We recall that the BRST transformations of all the fields are given by
\begin{align}\label{BRST1}
sA_{\mu }^{a} &=-\left( D_{\mu }c\right) ^{a}\,, & sc^{a} &=\frac{1}{2}gf^{abc}c^{b}c^{c}\,,   \nonumber \\
s\overline{c}^{a} &=b^{a}\,,&   sb^{a}&=0\,,  \nonumber \\
s\varphi _{i}^{a} &=\omega _{i}^{a}\,,&s\omega _{i}^{a}&=0\,,\nonumber \\
s\overline{\omega}_{i}^{a} &=\overline{\varphi }_{i}^{a}\,,& s \overline{\varphi }_{i}^{a}&=0\,.
\end{align}
The existence of this explicit breaking can be easily checked by releasing the BRST transformation $s$ onto the action $S_\GZ$,
\begin{eqnarray}
s S_\GZ &=&g \gamma^2 \int \d^d x f^{abc} \left( A^a_{\mu} \omega^{bc}_\mu -
 \left(D_{\mu}^{am} c^m\right)\left( \overline{\varphi}^{bc}_\mu + \varphi^{bc}_{\mu}\right)  \right)\,.
\end{eqnarray}
We refer to \cite{Dudal:2008sp} for more details concerning this breaking. \\
\\
In order to discuss the renormalizability of $S_\GZ$, we treat the breaking as a composite operator to be introduced into the action by means of a suitable set of external sources. This procedure can be done in a BRST invariant way, by embedding $S_\GZ$ into a larger action, namely
\begin{eqnarray}\label{brstinvariant}
\Sigma_\GZ &=& S_{\YM} + S_{\gf} + S_0 + S_\s \,,
\end{eqnarray}
whereby
\begin{eqnarray}
S_\s &=& s\int \d^d x\left( -U_\mu^{ai} D_\mu^{ab} \varphi_i^b - V_\mu^{ai} D_{\mu}^{ab} \overline \omega_i^{ab} - U_\mu^{ai} V_\mu^{ai} \right)\nonumber\\
&=& \int \d^d x \left( -M_\mu^{ai}  D_\mu^{ab} \varphi_i^b - gf^{abc} U_\mu^{ai}   D^{bd}_\mu c^d  \varphi_i^c + U_\mu^{ai}  D_\mu^{ab} \omega_i^b \right.
\nonumber \\
&& \left.  - N_\mu^{ai}  D_\mu^{ab} \overline \omega_i^b - V_\mu^{ai}  D_\mu^{ab} \overline \varphi_i^b + gf^{abc} V_\mu^{ai} D_\mu^{bd} c^d \overline \omega_i^c - M_\mu^{ai} V_\mu^{ai}+U_\mu^{ai} N_\mu^{ai} \right) \,.
\end{eqnarray}
We have introduced 4 new sources $U_\mu^{ai}$, $V_\mu^{ai}$, $M_\mu^{ai}$ and $N_\mu^{ai}$ with the following BRST transformations,
and
\begin{align}\label{BRST2}
sU_{\mu }^{ai} &= M_{\mu }^{ai}\,, & sM_{\mu }^{ai}&=0\,,  \nonumber \\
sV_{\mu }^{ai} &= N_{\mu }^{ai}\,, & sN_{\mu }^{ai}&=0\,.
\end{align}
This embedding into a larger action is necessary for the algebraic proof of the renormalizability as this heavily relies on having a BRST symmetry.  Replacing the sources with their physical values in the end, returns the Gribov-Zwanziger action,
\begin{eqnarray}\label{physlimit}
&& \left. U_\mu^{ai}\right|_{\phys} = \left. N_\mu^{ai}\right|_{\phys} = 0 \,,\\
&& \left. M_{\mu \nu }^{ab}\right|_{\phys}= \left.V_{\mu \nu}^{ab}\right|_{\phys}=\gamma^2 \delta ^{ab}\delta _{\mu \nu } \,,
\end{eqnarray}
as one can easily check.

\subsection{The Refined Gribov-Zwanziger action}
Let us explain the origin of the Refined Gribov-Zwanziger action. In the original Gribov-Zwanziger framework in 4 dimensions, one obtains an infrared suppressed, positivity violating gluon propagator which tends towards zero for zero momentum and an infrared enhanced ghost propagator. This behavior of the gluon and the ghost propagator stemming from the action $S_\GZ$ seemed to be in agreement with the lattice results for a long time. Until more recently, the authors of \cite{Cucchieri:2007md} discovered a completely different behavior of the propagators in the deep infrared working on larger lattices. Now the ghost propagator no longer seems to be enhanced and the gluon propagator reaches a finite value at zero momentum. Since the publication of \cite{Cucchieri:2007md}, more lattice data have confirmed these striking results \cite{Bogolubsky,Cucchieri2,Cucchieri3,Bornyakov,Bogolubsky2,Maas:2008ri}. Therefore, the Gribov-Zwanziger framework appeared to be in disagreement with these newest lattice data. However, in \cite{Dudal:2007cw,Dudal:2008sp}, we have shown that it is still possible to obtain results with the help of the Gribov-Zwanziger action which are in qualitative concordance with these new lattice data by taking into account the dynamics of the fields $(\overline \varphi_\mu^{ac}$, $\varphi_\mu^{ac} $, $\overline \omega_\mu^{ac}$,$\omega_\mu^{ac}$). This gives rise to additional non-perturbative effects within the Gribov-Zwanziger framework as, for instance,  the dimension two condensate $\braket{\overline \varphi_\mu^{ac}\varphi_\mu^{ac}-\overline \omega_\mu^{ac} \omega_\mu^{ac}}$, which has been found \cite{Dudal:2007cw,Dudal:2008sp} to be proportional to $\gamma^2$. It is apparent that the dynamics of these extra fields is highly entangled to the existence of the horizon. Therefore, we have refined the Gribov-Zwanziger action by explicitly adding the operator $\overline \varphi_\mu^{ac}\varphi_\mu^{ac}-\overline \omega_\mu^{ac} \omega_\mu^{ac}$  from the start, while preserving the renormalizability of the theory.\\
\\
The Refined Gribov-Zwanziger action is thus given by
\begin{eqnarray}\label{RGZ}
S_\RGZ &=& S_\GZ + S_{\overline{\varphi} \varphi} + S_{\en}\,,
\end{eqnarray}
whereby
\begin{eqnarray}\label{overeenstemming}
S_{\overline{\varphi} \varphi} &=& - M^2 \int \d^d x  \left( \overline \varphi^a_i \varphi^a_i - \overline \omega^a_i \omega^a_i \right)  \,, \nonumber\\
S_{\en} &=&   2 \frac{d (N^2 -1)}{\sqrt{2 g^2 N}}  \int \d^d x\ \varsigma \ \gamma^2 M^2 \,.
\end{eqnarray}
We have introduced a new parameter $\varsigma$ and a new mass $M^2$. The second term $S_\en$ is a constant term, which is comparable with the term $-\gamma ^{2}  \int \d^d x d \left(N^{2}-1\right)  \gamma^2 $ in the original Gribov-Zwanziger formulation \eqref{SGZ}. This term will allow us to remain inside the Gribov region $\Omega$. For more details on this construction, we refer the reader to \cite{Dudal:2008sp}.

\section{The (Refined) Gribov-Zwanziger action with the inclusion of the scalar glueball operator}
\subsection{Generalities}\label{generalities}
The most natural way to study the lightest scalar glueball is by determining the correlator\footnote{At least, this is our starting point. Later, we shall determine a renormalization group invariant $\mathcal R$ containing $F^2_{\mu\nu}$, so we can calculate $\Braket{\mathcal R(x) \mathcal R(y)}$.} $\Braket{\frac{F^2(x) }{4} \frac{F^2(y)}{4}}$.  This correlator can be obtained by adding the operator $F_{\mu\nu}^2 /4$ to the (Refined) Gribov-Zwanziger action by coupling it to a source $q(x)$. In this fashion, we obtain the correlator as follows,
 \begin{eqnarray}
\left[{\frac{\delta}{\delta q(y)}  \frac{\delta}{\delta q(x)}}
Z^c\right]_{q=0} &=& \Braket{ \frac{F^2(x)}{4} \frac{F^2(y)}{4}}\,,
\end{eqnarray}
with $Z^c$ the generator of connected Green functions. In \cite{Dudal:2008tg} we have studied the glueball operator in the standard Yang-Mills theory, supplemented with the Landau gauge fixing. The framework we have set up for pure Yang-Mills theories,  can be now extended to the more complex case of the Gribov-Zwanziger action, which is our current goal.\\
\\
Unfortunately, simply adding $F_{\mu\nu}^2$ to the action turns out to be too naive. In \cite{Dudal:2008tg}, we have seen that the 4 dimensional operator $F_{\mu\nu}^2$ mixes with other 4 dimensional operators in $d=4$, in agreement with the general theory concerning the renormalization of gauge invariant operators \cite{KlubergStern:1974rs,Joglekar:1975nu,Henneaux:1993jn}.\\\\Obviously, we also expect a similar mixing in the Gribov-Zwanziger framework. As outlined in \cite{Dudal:2008tg,Collins:1984xc,Collins:1994ee}, we can distinguish between 3 different classes of dimension 4 operators. The first class $C_1$ is the set of the gauge invariant operators, for example $F_{\mu\nu}^2$. The cohomology of the nilpotent BRST symmetry generator $s$ allows to identify the $C_1$ operators $\mathcal F$ as those which can be written as $s \mathcal F= 0$, but also $\mathcal F \not= s(\ldots)$. The second class $C_2$ are the BRST exact operators, which are trivially BRST invariant due to the nilpotency of the BRST operator. Thus $\mathcal E \in C_2$ if and only if $ \mathcal E = s (\ldots)$. The third class $C_3$ contains operators which vanish when the equations of motion are invoked. One can then argue that the mixing matrix of these operators must be upper triangular,
\begin{eqnarray}\label{upper}
\left( \begin{array}{c} \mathcal F_0 \\ \mathcal E_0 \\ \mathcal H_0 \end{array} \right) &= & \left( \begin{array}{ccc} Z_{\mathcal F\mathcal F}& Z_{\mathcal F\mathcal E}  & Z_{\mathcal F \mathcal H}\\ 0  &Z_{\mathcal E\mathcal E} & Z_{\mathcal E \mathcal H}   \\ 0 &
0& Z_{\mathcal H \mathcal H}          \end{array} \right) \left(
\begin{array}{c} \mathcal F \\ \mathcal E \\ \mathcal H \end{array}
\right)\,.
\end{eqnarray}
This particular behavior of the mixing of the various class of operators can be easily understood \cite{Collins:1984xc,Collins:1994ee}. Bare $C_2$ operators cannot receive contributions from gauge invariant $C_1$ operators: matrix elements of a bare BRST exact operator $\mathcal{E}$ between physical states are zero. But, if there would be a renormalized gauge invariant $C_1$ contribution in the expansion of $\mathcal{E}$, then there  would be room for a nonvanishing contribution, which is of course a contradiction. Likewise, any $C_3$ operator vanishes upon using the equations of motion, while $C_1$- and a $C_2$ operators in general do not, hence a $C_3$ operator will not receive corrections from the other type of operators.\\
\\
In \cite{Dudal:2008tg}, we have strictly proven in an algebraic fashion the upper triangular form of the mixing matrix for the operator $F_{\mu\nu}^2$, just by using the Ward identities of the action. In particular, we have proven that the following action is renormalizable for ordinary Yang-Mills gauge theories in the Landau gauge,
\begin{eqnarray}\label{glueballYM}
\Sigma_\glueYM &=&S_{\YM} + \int \d^d x\,\left( b^a \partial_\mu A_\mu^a + \overline c^a \p_\mu D_\mu^{ab} c^b \right) +  \int \d^d x q  \underbrace{\frac{1} {4} F_{\mu\nu}^2}_{\in C_1}  + \int \d^d x  \lambda \p_\mu \overline c^a A_\mu^a + \int \d^d x \eta \underbrace{\left(  \p_\mu b^a  A_\mu^a + \p_\mu\overline c^a D_\mu^{ab} c^b \right)}_{\in C_2}\nonumber\\
&&+  \int \d^d  x \alpha  \underbrace{ A_\mu^a\frac{\delta ( S_{\YM} + S_\gf) }{ \delta A_\mu^a}  }_{\in C_3} \,,
\end{eqnarray}
whereby we see the three different classes of operators arising. We have introduced three new sources: the doublet ($\lambda$,$\eta$) with $s\eta = \lambda$ and the color singlet $\alpha$. The term $\left(  \p_\mu b^a  A_\mu^a + \p_\mu\overline c^a D_\mu^{ab} c^b \right)$ is indeed an element belonging to the second class $C_2$, as we can rewrite it as $s (\p_\mu \overline c^a A^a_\mu)$. In \cite{Dudal:2008tg}, we have introduced the last term through a shift of the gluon field $A_\mu^a \rightarrow A_\mu^a + \alpha A_\mu^a$.

\subsection{Inclusion of the glueball operator in the Gribov-Zwanziger action}
With the mixing of the 4 dimensional operators in mind, we can propose an enlarged Gribov-Zwanziger action containing the glueball operator $F^2_{\mu\nu}$. This action will turn out to be renormalizable. For this, we can make two observations. Firstly, the limit, $\{\varphi, \overline \varphi, \omega, \overline \omega, U, V, N, M \} \to 0$, has to lead to our original Yang-Mills action $\Sigma_\glueYM$ with the addition of the glueball terms given by equation \eqref{glueballYM}. Secondly, setting all the terms related to the glueball term $q F^2$ equal to zero, we should recover the Gribov-Zwanziger action $\Sigma_\GZ$ in equation \eqref{brstinvariant}. Therefore, we propose the following starting action:
\begin{eqnarray}\label{glueballaction}
 \Sigma_\glue &=& \Sigma_{\GZ} + \int \d^d x \ q F_{\mu\nu}^a F_{\mu\nu}^a + \int \d^d x s \left( \eta  \left[  \p_\mu \overline c^a A_\mu^a + \p \overline \omega \p \varphi + g f_{akb} \p \overline \omega^a A^k \varphi^b + U^a D^{ab} \varphi^b + V^a D^{ab}\overline \omega^b + UV \right]\right) \nonumber\\
&=& \Sigma_{\mathrm{GZ}} + \int \d^d x \ q F_{\mu\nu}^a F_{\mu\nu}^a +  \int \d^d x (\lambda \left[  \p_\mu \overline c^a A_\mu^a + \p \overline \omega \p \varphi + g f_{akb} \p \overline \omega^a A^k \varphi^b + U^a D^{ab} \varphi^b + V^a D^{ab}\overline \omega^b + UV \right]  \nonumber\\
&&+ \eta \Bigl[ \p_\mu b^a A_\mu^a + \p_\mu \overline c^a D^{ab}_{\mu} c^b + \p \overline \varphi \p \varphi - \p \overline \omega \p \omega + g f_{akb} \p \overline \varphi^a A^k \varphi^b  + g f_{akb} \p \overline \omega^a D^{kd} c^d \varphi^b -  g f_{akb} \p \overline \omega^a A^k \omega^b \nonumber\\
 &&+M_\mu^{ai} D_\mu^{ab} \varphi_i^b + g U_\mu^{ai} f^{abc}  D_\mu^{ab}c^b \varphi_i^c - U_\mu^{ai}  D_\mu^{ab}\omega_i^b  + N_\mu^{ai} D_\mu^{ab} \overline \omega_i^b -gV_\mu^{ai} f^{abc} D_\mu^{bd}c^d \overline \omega_i^c + V_\mu^{ai}  D_\mu^{ab} \overline \varphi_i^b \nonumber\\
  &&+M_\mu^{ai} V_\mu^{ai}-U_\mu^{ai} N_\mu^{ai}\Bigr]\,.
\end{eqnarray}
Indeed, upon taking the limit $\{\varphi, \overline \varphi, \omega, \overline \omega, U, V, N, M \} \to 0$, we recover the Yang-Mills action\footnote{The term proportional to the equations of motion will be introduced later.} \eqref{glueballYM} and setting all sources equal to zero ($q$, $\eta$, $\lambda$) $\to 0$, we find our original Gribov-Zwanziger action back, see equation \eqref{brstinvariant}. Notice that in principle, we could have taken other possible starting actions which also enjoy these two correct limits. We could have tried to couple different sources to the different BRST exact terms instead of employing only one source $\eta$. However, this would not lead to a renormalizable action, while the action \eqref{glueballaction} does turn out to be renormalizable, as we shall prove.\\
\\
We shall now try to establish the renormalizability of \eqref{glueballaction} by using the algebraic renormalization formalism \cite{Piguet:1995er}. \\
\\
The first step is to introduce two auxiliary terms necessary for the process of renormalization. Firstly, we add an additional external term $S_{\ext,1}$ to the action,
\begin{eqnarray}\label{Sext1}
S_{\ext,1}&=&\int \d^d x \left( -K_\mu^a  D_\mu^{ab}  c^b + \frac{1}{2} g L^a f^{abc} c^b c^c  \right) \,,
\end{eqnarray}
which is needed to define the nonlinear BRST transformations of the gauge field $A_\mu^a$ and of the ghost field $c^a$. $K_\mu^a$ and $L^a$ are two new BRST invariant sources which shall be set equal to zero in the end,
\begin{align}
\left. K_\mu^a \right|_\phys &= 0\,, &  \left. L^a \right|_\phys &= 0\,.
\end{align}
Therefore, these sources can be seen as two auxiliary sources which do not change the physics of the theory. Secondly, we also introduce the following external term,
\begin{eqnarray}\label{Sext2}
S_{\ext,2} &=& \int \d^d x s(X_i A_\mu^a \p \overline \omega^a_i) ~=~ \int \d^d x Y_i  A_\mu^a \p \overline \omega^a_i - \int \d^d x \left( X_i D^{ab}_\mu c^b \p_\mu \overline \omega^a_i + X_i A^a_\mu \p_\mu \overline \varphi^a_i \right)\,,
\end{eqnarray}
whereby $(X_i, Y_i)$ is a new doublet of sources, i.e.~$s X_i = Y_i$. This additional term is necessary in order to have a sufficient powerful set of Ward identities. Without this term, two Ward identities of the original Gribov-Zwanziger action would be broken which are absolutely indispensable for the proof a the renormalization of the action (see Ward identity 8. and 9. in the list below). Again, in the end, we shall set
\begin{align}
\left. X_i \right|_\phys &= 0\,, &  \left. Y_i \right|_\phys &= 0\,,
\end{align}
We shall thus continue the analysis with the following action
\begin{eqnarray}
\Sigma &=& \Sigma_\glue + S_{\ext,1} + S_{\ext,2}\,.
\end{eqnarray}
\\
The second step is to search for all the Ward identities obeyed by the classical action $\Sigma$. Doing so, we find the following list of identities:
\begin{enumerate}
\item  The Slavnov-Taylor idenitity:
\begin{equation}
\mathcal{S}(\Sigma ) = 0  \;,
\end{equation}
where
\begin{equation}
\mathcal{S}(\Sigma ) ~=~\int \d^dx \left( \frac{\delta \Sigma}{\delta K_{\mu}^{a}}\frac{\delta \Sigma }{\delta A_{\mu}^{a}}+\frac{\delta \Sigma }{\delta L^{a}}\frac{\delta \Sigma
}{\delta c^{a}}+b^{a}\frac{\delta \Sigma}{\delta \overline{c}^{a}}+\overline{\varphi }_{i}^{a}\frac{\delta \Sigma }{\delta \overline{\omega }_{i}^{a}}+\omega _{i}^{a}\frac{\delta \Sigma }{\delta \varphi _{i}^{a}}+M_{\mu }^{ai}\frac{\delta \Sigma}{\delta U_{\mu}^{ai}}+N_{\mu }^{ai}\frac{\delta \Sigma }{\delta V_{\mu }^{ai}}+ \lambda \frac{\delta \Sigma }{\delta \eta} +  Y_i \frac{\delta \Sigma }{\delta X_i} \right) \,.
\end{equation}
This identity is a functional translation of the BRST invariance $s$.
\item The $U(f)$ invariance:
\begin{eqnarray}\label{wardid1}
U_{ij} \Sigma &=&0\,,
\end{eqnarray}
with
\begin{eqnarray}\label{Uij}
U_{ij}&=&\int \d^dx\left( \varphi_{i}^{a}\frac{\delta }{\delta \varphi _{j}^{a}}-\overline{\varphi}_{j}^{a}\frac{\delta }{\delta \overline{\varphi}_{i}^{a}}+\omega _{i}^{a}\frac{\delta }{\delta \omega _{j}^{a}}-\overline{\omega }_{j}^{a}\frac{\delta }{\delta \overline{\omega }_{i}^{a}}-  M^{aj}_{\mu} \frac{\delta}{\delta M^{ai}_{\mu}} -U^{aj}_{\mu}\frac{\delta}{\delta U^{ai}_{\mu}}   \right. \nonumber\\
&&\left. \hspace{1cm} + N^{ai}_{\mu}\frac{\delta}{\delta N^{aj}_{\mu}} +V^{ai}_{\mu}\frac{\delta}{\delta V^{aj}_{\mu}}  +Y^{i}\frac{\delta}{\delta Y^{j}} +X^{i}\frac{\delta}{\delta X^{j}}  \right)\,.
\end{eqnarray}
Using $Q_{f}=U_{ii}$, we can associate an extra quantum number to the $i$-valued fields and sources. One can find all quantum numbers in TABLE \ref{tabel3} and TABLE \ref{tabel4}.
\item The Landau gauge condition:
\begin{eqnarray}
\frac{\delta \Sigma}{\delta b^{a}}&=&\p_\mu A_\mu^a -\p_\mu(\eta A_\mu^a)\,.
\end{eqnarray}
\item The modified antighost equation :
\begin{eqnarray}
\frac{\delta \Sigma}{\delta \overline c^{a}}+\p_\mu\frac{\delta \Sigma}{\delta K_{\mu}^a}-\p_\mu\left( \eta\frac{\delta \Sigma}{\delta K_\mu^a } \right)&=& \p (\lambda A) \,.
\end{eqnarray}
\item The ghost Ward identity:
\begin{equation}
\mathcal{G}^{a}\Sigma =\Delta _{\mathrm{cl}}^{a}\,,
\end{equation}
with
\begin{eqnarray}
\mathcal{G}^{a} &=&\int \d^dx\left( \frac{\delta }{\delta c^{a}}+gf^{abc}\left( \overline{c}^{b}\frac{\delta }{\delta b^{c}}+\varphi _{i}^{b}\frac{\delta }{\delta \omega _{i}^{c}}+\overline{\omega }_{i}^{b}\frac{\delta }{\delta \overline{\varphi }_{i}^{c}}+V_{\mu }^{bi}\frac{\delta }{\delta N_{\mu }^{ci}}+U_{\mu }^{bi}\frac{\delta }{\delta M_{\mu }^{ci}}\right) \right) \,.  \nonumber \\
&&
\end{eqnarray}
\item\label{lincon} Two linearly broken local constraints:
\begin{eqnarray}\label{lincon2}
&&\frac{\delta\Sigma}{\delta \overline\varphi^{ai}}+\p_\mu\frac{\delta\Sigma}{\delta M_\mu^{ai}}=gf^{abc}A_\mu^b V_\mu^{ci}-\eta gf^{abc}A_{\mu }^{b}V_{\mu}^{ci}-\p_\mu ( X_i A^a_\mu) \,, \nonumber\\
&&\frac{\delta \Sigma}{\delta\omega^{ai}}+\partial_{\mu}\frac{\delta\Sigma}{\delta N_{\mu}^{ai}}-gf^{abc}\overline{\omega}^{bi}\frac{\delta\Sigma}{\delta b^{c}}=gf^{abc}A_{\mu}^{b}U_{\mu}^{ci}  - \eta gf^{abc}A_{\mu}^{b}U_{\mu}^{ci} \,.
\end{eqnarray}
\item  The exact $\mathcal{R}_{ij}$ invariance:
\begin{equation}\label{wardander}
\mathcal{R}_{ij}\Sigma =0\,,
\end{equation}
with
\begin{equation*}
\mathcal{R}_{ij}=\int \d^dx\left( \varphi _{i}^{a}\frac{\delta}{\delta\omega _{j}^{a}}-\overline{\omega }_{j}^{a}\frac{\delta }{\delta \overline{\varphi }_{i}^{a}}+V_{\mu }^{ai}\frac{\delta }{\delta N_{\mu}^{aj}}-U_{\mu }^{aj}\frac{\delta }{\delta M_{\mu }^{ai}} - X^i \frac{\delta}{\delta Y^j}\right) \,.
\end{equation*}
\item An extra integrated Ward identity:
\begin{eqnarray}\label{broken1}
\int \d^d x\left( \frac{\delta }{\delta \lambda} - \eta \frac{\delta }{\delta \lambda} + \overline c^a \frac{\delta }{\delta b^a}  + U_\mu^{ai} \frac{\delta }{\delta M_\mu^{ai}}  + \overline \omega_i^a  \frac{\delta }{\delta \overline \varphi_i^a}  - X_i  \frac{\delta }{\delta Y_i}    \right) \Sigma&=& 0\,,
\end{eqnarray}
which expresses in functional form the BRST exactness of the operator coupled to $\lambda$.
\item The integrated Ward Identity:
\begin{eqnarray}\label{broken2}
\int \d^d x \left( c^a \frac{\delta }{ \delta \omega^{a i} } + \overline{\omega}^{a i}  \frac{\delta }{ \delta \overline{c}^a }  + U^{ai}_\mu  \frac{\delta }{ \delta K^{a}_\mu} -\eta U^{ai}_\mu  \frac{\delta }{ \delta K^{a}_\mu} - \lambda \frac{\delta }{ \delta Y_i}  \right) \Sigma = 0\,.
\end{eqnarray}
\item The $X$-and $Y$-Ward identities:
\begin{eqnarray} \label{wardideinde}
\int \d^d x \left[ (1 - \eta) \frac{\delta}{ \delta X^i} - \lambda \frac{\delta }{ \delta Y^i} + \overline \omega^{a}_i \frac{\delta }{ \delta \overline c^a}   + \overline \varphi_i^a \frac{\delta }{ \delta b^a} \right] \Sigma &=& 0 \,,\nonumber\\
\int \d^d x \left[ (1 - \eta) \frac{\delta}{ \delta Y^i} + \overline \omega_i^a \frac{\delta }{ \delta b^a} \right] \Sigma &=& 0\,.
\end{eqnarray}
\end{enumerate}
\begin{table}[H]
  \centering
        \begin{tabular}{|c|c|c|c|c|c|c|c|c|}
        \hline
        & $A_{\mu }^{a}$ & $c^{a}$ & $\overline{c}^{a}$ & $b^{a}$ & $\varphi_{i}^{a} $ & $\overline{\varphi }_{i}^{a}$ &            $\omega _{i}^{a}$ & $\overline{\omega }_{i}^{a}$ \\
        \hline
        \hline
        \textrm{dimension} & $1$ & $0$ &$2$ & $2$ & $1$ & $1$ & $1$ & $1$ \\
        \hline
        $\mathrm{ghost\, number}$ & $0$ & $1$ & $-1$ & $0$ & $0$ & $0$ & $1$ & $-1$ \\
        \hline
        $Q_{f}\textrm{-charge}$ & $0$ & $0$ & $0$ & $0$ & $1$ & $-1$& $1$ & $-1$\\
        \hline
        \end{tabular}
        \caption{Quantum numbers of the fields.}\label{tabel3}
        \end{table}
        \begin{table}[H]
    \centering
    \begin{tabular}{|c|c|c|c|c|c|c|c|c|c|c|c|}
        \hline
        &$U_{\mu}^{ai}$&$M_{\mu }^{ai}$&$N_{\mu }^{ai}$&$V_{\mu }^{ai}$&$K_{\mu }^{a}$&$L^{a}$& $q$&  $\eta$& $\lambda$ & $X^i$ & $Y^i$ \\
        \hline
        \hline
        \textrm{dimension} & $2$ & $2$ & $2$ &$2$  & $3$ & $4$ & $0$ & $0$ & $0$ &1&1 \\
        \hline
        $\mathrm{ghost\, number}$ & $-1$& $0$ & $1$ & $0$ & $-1$ & $-2$ & 0 & 0 & 1 & 0 & 1 \\
        \hline
        $Q_{f}\textrm{-charge}$ & $-1$ & $-1$ & $1$ & $1$ & $0$ & $0$  & 0 & 0&0 & 1 & 1 \\
        \hline
        \end{tabular}
        \caption{Quantum numbers of the sources.}\label{tabel4}
\end{table}

Let us stress here that it is of paramount importance to have a good set of Ward identities to start from. For the construction of the action $\Sigma$, one should keep in mind the limits to the ordinary Gribov-Zwanziger case and to the Yang-Mills action with the inclusion of the glueball term. It is logical that an identity which plays a crucial role in one of the two limit cases, should not be broken by the action $\Sigma$, as $\Sigma$ can be seen as an enlargement of the two limit cases. This is the reason why we have introduced $S_{\ext,2}$. Without the auxiliary sources $X_i$ and $Y_i$, the extra integrated Ward identity \eqref{broken1} and the integrated Ward identity \eqref{broken2} are broken, and without these two identities one cannot prove the renormalizability of the action in an algebraic way. Let us also mention that in the ordinary Gribov-Zwanziger case, we have two extra linearly broken constraints, belonging to the set of Ward identities in equation \eqref{lincon2}. However, it is not a problem that these two identities are broken, as the other two linearly broken constraints in equation \eqref{lincon2} turn out to be equivalent at the level of the algebraic renormalization, namely: they have the same effect on the counterterm. \\
\\
Subsequently, we are ready to turn to quantum level. The third step is to characterize the most general integrated local counterterm $\Sigma^c$ which can be freely added to all orders of perturbation theory. $\Sigma^c$ is however restricted due to the existence of the Ward identities. Let us investigate these restrictions a bit closer. The classical action changes under quantum corrections according to
\begin{eqnarray}\label{deformed}
    \Sigma \rightarrow \Sigma + h \Sigma^c\,,
\end{eqnarray}
whereby $h$ is the perturbation parameter. Demanding that the perturbed action $ (\Sigma + h \Sigma^c) $ fulfills the same set of Ward identities obeyed by $\Sigma$, see \cite{Piguet:1995er}, it follows that the counterterm $\Sigma^c$ is constrained by:
\begin{enumerate}
\item  The linearized Slavnov-Taylor identity:
\begin{equation}\label{ST}
\mathcal{B}_{\Sigma }\Sigma^c=0\,,
\end{equation}
where $\mathcal{B}_{\Sigma }$ is the nilpotent linearized Slavnov-Taylor operator,
\begin{eqnarray*}
\mathcal{B}_{\Sigma} &=&\int \d^dx\left( \frac{\delta \Sigma}{\delta K_{\mu }^{a}}\frac{\delta }{\delta A_{\mu }^{a}}+\frac{\delta \Sigma }{\delta A_{\mu }^{a}}\frac{\delta }{\delta K_{\mu }^{a}}+\frac{\delta\Sigma }{\delta L^{a}}\frac{\delta }{\delta c^{a}}+\frac{\delta\Sigma }{\delta c^{a}}\frac{\delta }{\delta L^{a}}+b^{a}\frac{\delta }{\delta \overline{c}^{a}}+\overline{\varphi}_{i}^{a}\frac{\delta }{\delta \overline{\omega }_{i}^{a}}+\omega_{i}^{a}\frac{\delta }{\delta \varphi_{i}^{a}}\right.\nonumber\\
&& \hspace{5cm}\left.+M_{\mu }^{ai}\frac{\delta }{\delta U_{\mu }^{ai}}+N_{\mu }^{ai}\frac{\delta }{\delta V_{\mu }^{ai}} + \lambda \frac{\delta }{\delta \eta} + Y^i \frac{\delta  }{\delta X_i} \right)\,,
\end{eqnarray*}
and
\begin{equation}
\mathcal{B}_{\Sigma }\mathcal{B}_{\Sigma }=0\,.
\end{equation}
\item The $U(f)$ invariance:
\begin{eqnarray}\label{cont1}
U_{ij} \Sigma^c &=&0 \,.
\end{eqnarray}
$U_{ij}$ is given in expression \eqref{Uij}.
\item The Landau gauge condition
\begin{eqnarray}
\frac{\delta \Sigma^c}{\delta b^{a}}&=&0\,.
\end{eqnarray}
\item The modified antighost equation:
\begin{eqnarray}
\frac{\delta \Sigma^c}{\delta \overline c^{a}}+\p_\mu\frac{\delta \Sigma^c}{\delta K_{\mu}^a}-\p_\mu\left( \eta\frac{\delta \Sigma^c}{\delta K_\mu^a } \right)&=&0 \,.
\end{eqnarray}
\item The ghost Ward identity:
\begin{equation}
\mathcal{G}^{a}\Sigma^c =0\,,
\end{equation}
with
\begin{eqnarray}
\mathcal{G}^{a} &=&\int \d^dx\left( \frac{\delta }{\delta c^{a}}+gf^{abc}\left( \overline{c}^{b}\frac{\delta }{\delta b^{c}}+\varphi _{i}^{b}\frac{\delta }{\delta \omega _{i}^{c}}+\overline{\omega }_{i}^{b}\frac{\delta }{\delta \overline{\varphi }_{i}^{c}}+V_{\mu }^{bi}\frac{\delta }{\delta N_{\mu }^{ci}}+U_{\mu }^{bi}\frac{\delta }{\delta M_{\mu }^{ci}}\right) \right) \,.
\end{eqnarray}
\item  The linearly broken local constraints:
\begin{eqnarray}
&&\frac{\delta \Sigma^c}{\delta \overline{\varphi }^{ai}}+\partial _{\mu }\frac{\delta \Sigma^c }{\delta M_{\mu }^{ai}}=0\,, \nonumber\\
&&\frac{\delta \Sigma^c}{\delta\omega^{ai}}+\partial_{\mu}\frac{\delta\Sigma^c}{\delta N_{\mu}^{ai}}-gf^{abc}\overline{\omega}^{bi}\frac{\delta\Sigma^c}{\delta b^{c}}=0 \,.
\end{eqnarray}
\item  The exact $\mathcal{R}_{ij}$ symmetry:
\begin{equation}\label{cont2}
\mathcal{R}_{ij}\Sigma^c =0\,.
\end{equation}
\item The extra integrated Ward identity:
\begin{eqnarray}
\int \d^d x\left( \frac{\delta }{\delta \lambda} - \eta \frac{\delta }{\delta \lambda} + \overline c^a \frac{\delta }{\delta b^a}  + U_\mu^{ai} \frac{\delta }{\delta M_\mu^{ai}}  + \overline \omega_i^a  \frac{\delta }{\delta \overline \varphi_i^a}  - X_i  \frac{\delta }{\delta Y_i}    \right) \Sigma^c&=& 0\,.
\end{eqnarray}
\item The integrated Ward Identity:
\begin{eqnarray}
\int \d^d x \left( c^a \frac{\delta }{ \delta \omega^{a i} } + \overline{\omega}^{a i}  \frac{\delta }{ \delta \overline{c}^a }  + U^{ai}_\mu  \frac{\delta }{ \delta K^{a}_\mu} -\eta U^{ai}_\mu  \frac{\delta }{ \delta K^{a}_\mu} - \lambda \frac{\delta }{ \delta Y_i}  \right) \Sigma^c = 0\,.
\end{eqnarray}
\item The $X$-and $Y$-Ward identities:
\begin{eqnarray}\label{cont3}
\int \d^d x \left[ (1 - \eta) \frac{\delta}{ \delta X^i} - \lambda \frac{\delta }{ \delta Y^i} + \overline \omega^{a}_i \frac{\delta }{ \delta \overline c^a}   + \overline \varphi_i^a \frac{\delta }{ \delta b^a} \right] \Sigma^c &=& 0 \,, \nonumber\\
\int \d^d x \left[ (1 - \eta) \frac{\delta}{ \delta Y^i} + \overline \omega_i^a \frac{\delta }{ \delta b^a} \right] \Sigma^c &=& 0\,.
\end{eqnarray}
\end{enumerate}
At this point, we are ready to determine the most general integrated local polynomial $\Sigma^c$ in the fields and external sources of dimension bounded by four and with zero ghost number, limited by the constraints \eqref{ST}--\eqref{cont3}. The linearized Slavnov-Taylor identity plays an important role in simplifying the form of the counterterm. Indeed, the counterterm can be parameterized as follows:
\begin{eqnarray}
\Sigma^c &=& \underbrace{\left(\mathcal{B}_{\Sigma}\textrm{ closed but not exact part}\right)}_{\Sigma^c_1} + \underbrace{\mathcal{B}_{\Sigma }\Delta^{-1}}_{\Sigma^c_2}\,,
\end{eqnarray}
whereby $\Sigma^c_1$ is a cohomologically non-trivial part while $\Sigma^c_2$ represents the
cohomologically trivial part. $\Delta^{-1}$ is the most general local polynomial with dimension 4 and ghost number $-1$. One
can prove that all fields and sources belonging to a doublet can only enter the cohomologically trivial part \cite{Piguet:1995er}. This
is exactly the reason why we have opted to introduce the source $\eta$, which is coupled to the BRST exact term, as part of a doublet. In this way, the source $\eta$ can only enter the trivial part, and turns out to be useful to explicitly prove the upper triangular form of the mixing matrix in equation \eqref{upper}. One can now check that the closed but not exact part is given by
\begin{eqnarray}
\Sigma^c_1 &=&a_0 S_{\YM} + b_0 \widehat S_{\YM}\,,
\end{eqnarray}
whereby
\begin{eqnarray}
\widehat S_{\YM} &=& \int \d^d x q \frac{1}{4} F_{\mu\nu}^a F_{\mu\nu}^a \,,
\end{eqnarray}
and the trivial part is given by the following rather lengthy expression:
\begin{eqnarray}\label{counterruw}
\Sigma^c_2 &=&    \mathcal{B}_\Sigma \int \d^d \!x\,   \biggl\{ \biggl[ a_{1}(K_{\mu}^{a}+\partial _{\mu} \overline{c}^{a})A_{\mu}^{a}+a_{2}\,L^{a}c^{a} +a_{3}U_{\mu i}^{a}\,\partial _{\mu }\varphi _{i}^{a} +a_{4}\,V_{\mu i}^{a}\,\partial _{\mu }\overline{\omega }_{i}^{a} +a_{5}\,\overline{\omega }_{i}^{a}\partial ^{2}\varphi _{i}^{a} \nonumber \\
&&+a_{6}\, \,U_{\mu i}^{a}V_{\mu i}^{a}+a_{7}\,gf^{abc}U_{\mu i}^{a}\,\varphi _{i}^{b}A_{\mu }^{c}+a_{8}\,gf^{abc}V_{\mu i}^{a}\,\overline{\omega }_{i}^{b}A_{\mu }^{c}+a_{9}\,gf^{abc}\overline{\omega }_{i}^{a}A_{\mu }^{c}\,\partial _{\mu }\varphi _{i}^{b} +a_{10}\,gf^{abc}\overline{\omega }_{i}^{a}(\partial _{\mu }A_{\mu}^{c})\varphi _{i}^{b}  \nonumber\\
&&+ a_{11} X^i \overline \omega^a_i \p A^a_\mu + a_{12} X^i \p\overline  \omega^a_i  A^a_\mu + a_{13} X^i \overline \varphi^a_i \overline c^a + a_{14}g f_{abc} X^i \omega^a_i \overline \omega^b_j \overline \omega^c_j +  a'_{14} g f_{abc} X^i \omega^a_j \overline \omega^b_i \overline \omega^c_j  \nonumber\\
&&  + a_{15} X^i \overline \omega^a_i b^a + a_{16} X^i U^{i a}_\mu A^a_\mu + a_{17} g f_{abc} X^i \overline \omega^a_i \varphi^b_j \overline \varphi_j^c +  a'_{17} g f_{abc} X^i \overline \omega^a_j \varphi^b_i \overline \varphi_j^c + + a_{17}^{\prime\prime} g f_{abc} X^i \overline \omega^a_j \varphi^b_j \overline \varphi_i^c \nonumber\\
&&+ a_{18} g f_{abc} X^i \overline \omega^a_i \overline c^b c^c +a_{19} X^i X^i \overline \varphi^a_j \overline \omega^a_j + a_{19}^\prime X^i X^j \overline \varphi^a_i \overline \omega^a_j + a_ {20}X^i Y^j \overline \omega^i_a \overline \omega^j_a + a_{21}g f_{abc} Y^i \overline \omega^a_i \overline \omega^b_j \varphi^c_j \nonumber\\
&&+ a'_{21}g f_{abc} Y^i \overline \omega^a_j \overline \omega^b_j \varphi^c_i  +  a_{22} Y^i \overline \omega_i^a \overline c^a \biggr]\nonumber\\
&& +q \biggl[ b_{1} (K_{\mu}^{a}+\partial _{\mu} \overline{c}^{a})A_{\mu}^{a} +  c_{1}  \overline{c}^{a} \partial _{\mu} A_{\mu}^{a}  +b_{2} L^{a}c^{a} +b_{3} U_{\mu i}^{a}\,\partial _{\mu }\varphi _{i}^{a}+ c_{3} \partial _{\mu }U_{\mu i}^{a}\varphi _{i}^{a} +b_{4} V_{\mu i}^{a}\,\partial _{\mu }\overline{\omega }_{i}^{a} +c_{4} \partial _{\mu } V_{\mu i}^{a}\,\overline{\omega }_{i}^{a}   \nonumber\\
&&+b_{5} \overline{\omega }_{i}^{a}\partial ^{2}\varphi _{i}^{a}+  c_{5} \partial_{\mu}\overline{\omega }_{i}^{a}\partial_{\mu}\varphi _{i}^{a}  + d_{5} \partial ^{2} \overline{\omega }_{i}^{a}\varphi _{i}^{a}  +b_{6} \,U_{\mu i}^{a}V_{\mu i}^{a}+b_{7} gf^{abc}U_{\mu i}^{a}\,\varphi _{i}^{b}A_{\mu }^{c}+  b_{8} gf^{abc}V_{\mu i}^{a}\,\overline{\omega }_{i}^{b}A_{\mu }^{c}\nonumber \\
&&+b_{9} gf^{abc}\overline{\omega }_{i}^{a}A_{\mu }^{c}\,\partial _{\mu }\varphi _{i}^{b} +c_{9} gf^{abc}\overline{\omega }_{i}^{a}(\partial _{\mu }A_{\mu}^{c})\varphi _{i}^{b} + d_{9} gf^{abc}\partial _{\mu } \overline{\omega }_{i}^{a}A_{\mu}^{c}\varphi _{i}^{b} +b_{10} X^i \overline \omega^a_i \p A^a_\mu + c_{10} X^i \p\overline  \omega^a_i  A^a_\mu \nonumber\\
&& + d_{10} \p X^i \overline  \omega^a_i  A^a_\mu+ b_{11} X^i \overline \varphi^a_i \overline c^a + b_{12}g f_{abc} X^i \omega^a_i \overline \omega^b_j \overline \omega^c_j + b_{12}'g f_{abc} X^i \omega^a_j \overline \omega^b_i \overline \omega^c_j + b_{13} X^i \overline \omega^a_i b^a + b_{14} X^i U^{i a}_\mu A^a_\mu\nonumber\\
&&  + b_{15} g f_{abc} X^i \overline \omega^a_i \varphi^b_j \overline \varphi_j^c +b_{15}' g f_{abc} X^i \overline \omega^a_j \varphi^b_i \overline \varphi_j^c + b_{15}'' g f_{abc} X^i \overline \omega^a_j \varphi^b_j \overline \varphi_i^c  + b_{16} g f_{abc} X^i \overline \omega^a_i \overline c^b c^c +b_{17} X^i X^i \overline \varphi^a_j \overline \omega^a_j \nonumber\\
&&+ b_{17}^\prime X^i X^j \overline \varphi^a_i \overline \omega^a_j + b_ {18}X^i Y^j \overline \omega^i_a \overline \omega^j_a + b_{19}g f_{abc} Y^i \overline \omega^a_i \overline \omega^b_j \varphi_j^c+ b_{19}^\prime g f_{abc} Y^i \overline \omega^a_j \overline \omega^b_j \varphi_i^c + b_{20} Y^i \overline \omega_i^a \overline c^a \biggr]\nonumber\\
&&+ \eta \biggl[ e_{1}  K_{\mu}^{a}A_{\mu}^{a}+ e_1'\partial _{\mu} \overline{c}^{a} A_{\mu}^{a} +  f_{1}  \overline{c}^{a} \partial _{\mu} A_{\mu}^{a}  +e_{2} L^{a}c^{a} +e_{3}  U_{\mu i}^{a}\,\partial _{\mu }\varphi _{i}^{a}+ f_{3} \partial _{\mu }U_{\mu i}^{a}\varphi _{i}^{a} +e_{4} V_{\mu i}^{a}\,\partial _{\mu }\overline{\omega }_{i}^{a} +f_{4}\partial _{\mu } V_{\mu i}^{a}\,\overline{\omega }_{i}^{a} \nonumber\\
&& +e_{5}\,\overline{\omega }_{i}^{a}\partial ^{2}\varphi _{i}^{a} +  f_{5} \partial_{\mu}\overline{\omega }_{i}^{a}\partial_{\mu}\varphi _{i}^{a}  + g_{5} \partial ^{2} \overline{\omega }_{i}^{a}\varphi _{i}^{a}  +e_{6} \,U_{\mu i}^{a}V_{\mu i}^{a}+e_{7} gf^{abc}U_{\mu i}^{a}\,\varphi _{i}^{b}A_{\mu }^{c} +e_8 gf^{abc}V_{\mu i}^{a}\,\overline{\omega }_{i}^{b}A_{\mu }^{c}\nonumber \\
&&+e_{9} gf^{abc}\overline{\omega }_{i}^{a}A_{\mu }^{c}\,\partial _{\mu }\varphi _{i}^{b} +f_{9} gf^{abc}\overline{\omega }_{i}^{a}(\partial _{\mu }A_{\mu}^{c})\varphi _{i}^{b} + g_{9} gf^{abc}\partial _{\mu } \overline{\omega }_{i}^{a}A_{\mu}^{c}\varphi _{i}^{b} +e_{10} X^i \overline \omega^a_i \p A^a_\mu + f_{10} X^i \p\overline  \omega^a_i  A^a_\mu \nonumber\\
&& + g_{10} \p X^i \overline  \omega^a_i  A^a_\mu  + e_{11} X^i \overline \varphi^a_i \overline c^a + e_{12}g f_{abc} X^i \omega^a_i \overline \omega^b_j \overline \omega^c_j +e_{12}'g f_{abc} X^i \omega^a_j \overline \omega^b_i \overline \omega^c_j   + e_{13} X^i \overline \omega^a_i b^a + e_{14} X^i U^{i a}_\mu A^a_\mu \nonumber\\
&&  + e_{15} g f_{abc} X^i \overline \omega^a_i \varphi_b^j \overline \varphi^j_c + e_{15}' g f_{abc} X^i \overline \omega^a_j \varphi^b_i \overline \varphi_j^c + e_{15}'' g f_{abc} X^i \overline \omega^a_j \varphi^b_j \overline \varphi_i^c  + e_{16} g f_{abc} X^i \overline \omega^a_i \overline c^b c^c +e_{17} X^i X^i \overline \varphi^a_j \overline \omega^a_j \nonumber\\
&&+ e_{17}^\prime X^i X^j \overline \varphi^a_i \overline \omega^a_j + e_ {18}X^i Y^j \overline \omega^i_a \overline \omega^j_a + e_{19}g f_{abc} Y^i \overline \omega^a_i \overline \omega^b_j \varphi^c_j + e_{19}'g f_{abc} Y^i \overline \omega^a_j \overline \omega^b_j \varphi^c_i + e_{20} Y^i \overline \omega_i^a \overline c^a \biggr]\nonumber\\
&& \lambda \biggl[  h_1 g f_{abc} X^i  \varphi^{aj} \overline \omega^b_i \overline \omega^c_j + h_1' g f_{abc} X^i  \varphi^{ai} \overline \omega^b_j \overline \omega^c_j + h_2 X^i \overline c^a \overline \omega^a_i   + h_3 \overline \omega^a_i \overline \omega^b_j \varphi^a_i \varphi^b_j + (\mbox{variants of }h_3) \biggr]  \biggr\}\,.
\end{eqnarray}
The coefficients $a_i$, $a'_i$, etc.~are a priori free parameters.\\
\\
As the attentive reader might have noticed, we did not include terms of the form $(q^2 \ldots)$, $(\eta^2 \ldots)$, $(q \eta \ldots)$, $(q^3 \ldots)$, $(\lambda q^2 \ldots)$ etc., into the counterterm. However, by just looking at the dimensionality, the ghost number and the constraints on the counterterm, one might conclude that certain terms of quadratic and higher order in the sources ($q$, $\eta$, $\lambda$) are perfectly allowed. One can imagine that an infinite tower of counterterms would then be generated and thence it would be impossible to prove the renormalizability of the action as  new divergences are always being generated, which cannot be absorbed in terms already present in the classical action. However, we can give a simple argument why one may omit this class of terms with the help of an example. Assume that we would introduce the following term of order $q^2$ in the action,
\begin{eqnarray}\label{term}
&\sim&  \int \d^d x q^2 \frac{F_{\mu\nu}^2}{4}\,.
\end{eqnarray}
Subsequently, when calculating the correlator, this term would give rise to an extra contact term contribution,
\begin{eqnarray}\label{argument}
\left[\frac{\delta}{\delta q(z)}\frac{\delta }{\delta q(y)}\int [\d \phi] \e^{-\Sigma} \right]_{q=0}&=& \underbrace{\Braket{\frac{F^{2}(z)}{4} \frac{F^{2}(y)}{4}}}_{\mathrm{term\ due\ to\ part\ in\ }q} + \underbrace{\delta(y-z) \Braket{ \frac{F^2(y)}{2}}}_{\mathrm{term\ due\ to\ part\ in\ }q^2} \,.
\end{eqnarray}
Eventually, we are only interested in the correlator for $z \not = y$ and therefore we can neglect the term \eqref{term} quadratic in the source $q$. Moreover, when studying the case $z =y$, one should also couple a source to the novel composite operator $F^4 \equiv F^2_{\mu\nu} F^2_{\alpha\beta}$, which is not in our current interest. We can repeat this argument for all the terms which are \textit{zero} in the physical limit. Therefore, this argument is not only valid for the dimensionless sources $q$, $\eta$ and $\lambda$, but also for the massive sources $K_\mu$, $L_\mu$, $X_i$, $Y_i$. Though, some care needs to be taken. Let us explain this again with an example. The modified antighost equation has the following form:
\begin{eqnarray}
\frac{\delta \Sigma^c}{\delta \overline c^{a}}+\p_\mu\frac{\delta \Sigma^c}{\delta K_{\mu}^a}-\p_\mu\left( \eta\frac{\delta \Sigma^c}{\delta K_\mu^a } \right)&=& 0\,.
\end{eqnarray}
In this case, due to the term $\p_\mu\frac{\delta \Sigma^c}{\delta K_{\mu}^a}$, one compares terms of quadratic order in the sources $\sim q K_\mu \ldots$, with terms of first order in the sources $\sim q \ldots$. This identity can never be fulfilled is one immediately omits all terms of quadratic order in $K_\mu^a$. Therefore, we have chosen to keep all the possible combinations of higher order in the massive sources in the counterterm \eqref{counterruw} as there are only a finite number of combinations, while keeping in mind the higher order combinations of the dimensionless sources. Only after imposing all the constraints, we can then safely neglect the terms quadratic in the sources.\\
\\
With the previous remark in mind, we can now impose all the constraints \eqref{cont1}-\eqref{cont3} on the counterterm, which is a very cumbersome job. We ultimately find
\begin{eqnarray}
\Sigma^c&=& a_{0}S_{YM} +  b_0 \widehat S_{YM} +
a_{1}\int \d^dx\Biggl(  A_{\mu}^{a}\frac{ \delta S_{YM}}{\delta A_{\mu }^{a}}+ A_{\mu}^{a}\frac{\delta \widehat{S}_{YM}}{\delta A_{\mu }^{a}}  + \p_\mu \overline{c}^a \p_\mu c^a + K_{\mu }^{a}\partial _{\mu }c^{a}  + M_\mu^{a i} \p_\mu \varphi_\mu^{ai} -  U_\mu^{a i} \p_\mu \omega_\mu^{ai} \nonumber\\
&&N_\mu^{a i} \p_\mu \overline{\omega}_\mu^{ai} +  V_\mu^{a i} \p_\mu \overline{\varphi}_\mu^{ai}  +  \p_\mu \overline{\varphi}^{a i} \p_\mu \varphi_\mu^{ai} +  \p_\mu \omega^{a i} \p_\mu \overline{\omega}_\mu^{ai} + V_\mu^{a i} M_\mu^{a i} - U_\mu^{a i}N_\mu^{a i} - g f_{abc} U_\mu^{ia} \varphi^{bi} \p_\mu c^c \nonumber\\
&&- g f_{abc} V_\mu^{ia} \overline{\omega}^{bi} \p_\mu c^c - g f_{abc} \p_{\mu} \overline{\omega}^a \varphi^{bi}  \p_\mu c^c \Biggr)  \nonumber\\
&&+ b_{1}\int \d^dx q \Biggl(  A_{\mu}^{a}\frac{ \delta S_{YM}}{\delta A_{\mu }^{a}}  + \p_\mu \overline{c}^a \p_\mu c^a + K_{\mu }^{a}\partial _{\mu }c^{a}  + M_\mu^{a i} \p_\mu \varphi_\mu^{ai} -  U_\mu^{a i} \p_\mu \omega_\mu^{ai} +  N_\mu^{a i} \p_\mu \overline{\omega}_\mu^{ai} +  V_\mu^{a i} \p_\mu \overline{\varphi}_\mu^{ai} \nonumber\\
&&+  \p_\mu \overline{\varphi}^{a i} \p_\mu \varphi_\mu^{ai} +  \p_\mu \omega^{a i} \p_\mu \overline{\omega}_\mu^{ai} + V_\mu^{a i} M_\mu^{a i} - U_\mu^{a i}N_\mu^{a i} - g f_{abc} U_\mu^{ia} \varphi^{bi} \p_\mu c^c - g f_{abc} V_\mu^{ia} \overline{\omega}^{bi} \p_\mu c^c - g f_{abc} \p_{\mu} \overline{\omega}^a \varphi^{bi}  \p_\mu c^c \Biggr) \nonumber\\
&&+ a_{1}\int \d^dx \eta \Biggl(  \p_\mu \overline{c}^a \p_\mu c^a  + M_\mu^{a i} \p_\mu \varphi_\mu^{ai} -  U_\mu^{a i} \p_\mu \omega_\mu^{ai} +  N_\mu^{a i} \p_\mu \overline{\omega}_\mu^{ai} +  V_\mu^{a i} \p_\mu \overline{\varphi}_\mu^{ai} \nonumber\\
&&+  \p_\mu \overline{\varphi}^{a i} \p_\mu \varphi_\mu^{ai} +  \p_\mu \omega^{a i} \p_\mu \overline{\omega}_\mu^{ai} + V_\mu^{a i} M_\mu^{a i} - U_\mu^{a i}N_\mu^{a i} - g f_{abc} U_\mu^{ia} \varphi^{bi} \p_\mu c^c - g f_{abc} V_\mu^{ia} \overline{\omega}^{bi} \p_\mu c^c - g f_{abc} \p_{\mu} \overline{\omega}^a \varphi^{bi}  \p_\mu c^c \Biggr) \nonumber\\
&& +  a_{1}\int \d^dx \lambda \Biggl( U_\mu^{a i} \p_\mu \varphi^{a i} +  V_\mu^{a i} \p_\mu \overline \omega^{a i} + \p_\mu \overline \omega^{ai} \p_\mu \varphi^{ai} + U_\mu^{a i} V_\mu^{a i}  \Biggr) - a_{1}\int \d^dx  \Biggl( X^i \p_\mu \overline \omega^{ai} \p_\mu c^a  \Biggr)\,.
\end{eqnarray}
Only now, we can discard the term $\sim q K_{\mu }^{a}\partial
_{\mu }c^{a}$ as it is of quadratic order in the sources. One
could argue that we can also neglect terms of higher order in
$U_\mu^{ai}$ and $N_\mu^{ai}$. However, both sources belong to a BRST doublet. Moreover, the corresponding partner
sources, $M_\mu^{ai}, V_\mu^{ai}$, acquire a nonzero value in the
physical limit, and it would be impossible to write the BRST exact
term in our starting action $\Sigma_\glue$ (see expression
\eqref{glueballaction}) as an $s$-variation when neglecting these
kind of terms. In summary, the expression
\begin{eqnarray}\label{countertermfinal}
\Sigma^c&=& a_{0}S_{YM} +  b_0 \widehat S_{YM} +
a_{1}\int \d^dx\Biggl(  A_{\mu}^{a}\frac{ \delta S_{YM}}{\delta A_{\mu }^{a}}+ A_{\mu}^{a}\frac{\delta \widehat{S}_{YM}}{\delta A_{\mu }^{a}}  + \p_\mu \overline{c}^a \p_\mu c^a + K_{\mu }^{a}\partial _{\mu }c^{a}  + M_\mu^{a i} \p_\mu \varphi_\mu^{ai} -  U_\mu^{a i} \p_\mu \omega_\mu^{ai} \nonumber\\
&&N_\mu^{a i} \p_\mu \overline{\omega}_\mu^{ai} +  V_\mu^{a i} \p_\mu \overline{\varphi}_\mu^{ai}  +  \p_\mu \overline{\varphi}^{a i} \p_\mu \varphi_\mu^{ai} +  \p_\mu \omega^{a i} \p_\mu \overline{\omega}_\mu^{ai} + V_\mu^{a i} M_\mu^{a i} - U_\mu^{a i}N_\mu^{a i} - g f_{abc} U_\mu^{ia} \varphi^{bi} \p_\mu c^c \nonumber\\
&&- g f_{abc} V_\mu^{ia} \overline{\omega}^{bi} \p_\mu c^c - g f_{abc} \p_{\mu} \overline{\omega}^a \varphi^{bi}  \p_\mu c^c \Biggr)  \nonumber\\
&&+ b_{1}\int \d^dx q \Biggl(  A_{\mu}^{a}\frac{ \delta S_{YM}}{\delta A_{\mu }^{a}}  + \p_\mu \overline{c}^a \p_\mu c^a  + M_\mu^{a i} \p_\mu \varphi_\mu^{ai} -  U_\mu^{a i} \p_\mu \omega_\mu^{ai} +  N_\mu^{a i} \p_\mu \overline{\omega}_\mu^{ai} +  V_\mu^{a i} \p_\mu \overline{\varphi}_\mu^{ai} \nonumber\\
&&+  \p_\mu \overline{\varphi}^{a i} \p_\mu \varphi_\mu^{ai} +  \p_\mu \omega^{a i} \p_\mu \overline{\omega}_\mu^{ai} + V_\mu^{a i} M_\mu^{a i} - U_\mu^{a i}N_\mu^{a i} - g f_{abc} U_\mu^{ia} \varphi^{bi} \p_\mu c^c - g f_{abc} V_\mu^{ia} \overline{\omega}^{bi} \p_\mu c^c - g f_{abc} \p_{\mu} \overline{\omega}^a \varphi^{bi}  \p_\mu c^c \Biggr) \nonumber\\
&&+ a_{1}\int \d^dx \eta \Biggl(  \p_\mu \overline{c}^a \p_\mu c^a  + M_\mu^{a i} \p_\mu \varphi_\mu^{ai} -  U_\mu^{a i} \p_\mu \omega_\mu^{ai} +  N_\mu^{a i} \p_\mu \overline{\omega}_\mu^{ai} +  V_\mu^{a i} \p_\mu \overline{\varphi}_\mu^{ai} \nonumber\\
&&+  \p_\mu \overline{\varphi}^{a i} \p_\mu \varphi_\mu^{ai} +  \p_\mu \omega^{a i} \p_\mu \overline{\omega}_\mu^{ai} + V_\mu^{a i} M_\mu^{a i} - U_\mu^{a i}N_\mu^{a i} - g f_{abc} U_\mu^{ia} \varphi^{bi} \p_\mu c^c - g f_{abc} V_\mu^{ia} \overline{\omega}^{bi} \p_\mu c^c - g f_{abc} \p_{\mu} \overline{\omega}^a \varphi^{bi}  \p_\mu c^c \Biggr) \nonumber\\
&& +  a_{1}\int \d^dx \lambda \Biggl( U_\mu^{a i} \p_\mu \varphi^{a i} +  V_\mu^{a i} \p_\mu \overline \omega^{a i} + \p_\mu \overline \omega^{ai} \p_\mu \varphi^{ai} + U_\mu^{a i} V_\mu^{a i}  \Biggr) - a_{1}\int \d^dx  \Biggl( X^i \p_\mu \overline \omega^{ai} \p_\mu c^a  \Biggr)\,,
\end{eqnarray}
gives the general counterterm compatible with all Ward identities. \\
\\
We still need to introduce the operators belonging to
the class $C_3$, which are related to the equations of motion, see
section \ref{generalities}. Therefore, the next step is to perform
a linear shift on the gluon field $A^a_\mu$ in the action $\Sigma$
\begin{eqnarray}
A^a_\mu \rightarrow A^a_\mu + \alpha A^a_\mu\,,
\end{eqnarray}
whereby $\alpha$ is a dimensionless new source. As this shift corresponds to a redefinition of the gluon field it has to be consistently done in the starting action as well as in the counterterm. Later on, we shall see that introducing the relevant gluon equation of motion operator through this shift, will allow us to uncover the finiteness of this kind of operator. Performing the shift in the classical action yields the following shifted action $\Sigma'$
\begin{eqnarray}\label{eindactie}
 \Sigma' &=& S_{\mathrm{YM}} + \int \d^d x\,\left( b^{a}\partial_\mu A_\mu^{a}+\overline{c}^{a}\partial _{\mu } D_{\mu}^{ab}c^b \right) +\int \d^dx \left( -K_{\mu }^{a}\left( D_{\mu }c\right) ^{a}+\frac{1}{2}gL^{a}f^{abc}c^{b}c^{c}\right)   \nonumber\\
&&+\int \d^d x\left( \overline{\varphi }_i^a \partial_{\nu}  D_\nu^{ab} \varphi_i^b - \overline{\omega}_{i}^{a}\partial_\nu D_\nu^{ab} \omega_i^b  -g \partial_\nu \overline \omega_i^a f^{abm}  D_\nu^{bd} c^d \varphi_i^m \right)\nonumber\\
&&+ \int \d^dx \Bigl( -M_{\mu }^{ai} D_\mu^{ab} \varphi_i^b - g U_\mu^{ai} f^{abc}  D_\mu^{bd} c^d \varphi_i^c + U_\mu^{ai}  D_\mu^{ab} \omega_i^b \nonumber\\
&&  - N_\mu^{ai} D_\mu^{ab} \overline \omega_i^b  - V_\mu^{ai} D_\mu^{ab} \overline{\varphi}_i^b + g V_\mu^{ai} f^{abc} D_\mu^{bd} c^d \overline \omega_i^c - M_{\mu }^{ai}V_{\mu }^{ai}+U_{\mu }^{ai}N_{\mu }^{ai}\Bigr) \nonumber\\
  && + \int \d^d x  q F_{\mu\nu}^a F_{\mu\nu}^a +  \int \d^d x \lambda \left[  \p_\mu \overline c^a A_\mu^a + \p \overline \omega \p \varphi + g f_{akb} \p \overline \omega^a A^k \varphi^b + U^a D^{ab} \varphi^b + V^a D^{ab}\overline \omega^b + UV \right]  \nonumber\\
&&+\int \d^d x  \eta \Bigl[\p_\mu b^a A_\mu^a + \p_\mu \overline c^a D^{ab}_{\mu} c^{b} + \p \overline \varphi \p \varphi - \p \overline \omega \p \omega + g f_{akb} \p \overline \varphi^a A^k \varphi^b  + g f_{akb} \p \overline \omega^a D^{kd} c^d \varphi^b -  g f_{akb} \p \overline \omega^a A^k \omega^b \nonumber\\
 &&+M_{\mu }^{ai}\left( D_{\mu }\varphi_{i}\right) ^{a}+gU_{\mu }^{ai}f^{abc}\left( D_{\mu }c\right)^{b}\varphi _{i}^{c}-U_{\mu }^{ai}\left( D_{\mu }\omega _{i}\right)^{a}  + N_{\mu }^{ai}\left( D_{\mu }\overline{\omega }_{i}\right)^{a}-gV_{\mu}^{ai}f^{abc}\left(D_{\mu }c\right)^{b}\overline{\omega}_{i}^{c} + V_{\mu }^{ai}\left( D_{\mu }\overline{\varphi}_{i}\right)^{a}\nonumber\\
 &&+M_{\mu }^{ai}V_{\mu }^{ai}-U_{\mu }^{ai}N_{\mu }^{ai}\Bigr] + \int \d^d x \left( Y_i  A_\mu^a \p \overline \omega^a_i - X_i D^{ab}_\mu c^b \p_\mu \overline \omega^a_i + X_i A^a_\mu \p_\mu \overline \varphi^a_i\right) \nonumber\\
&&+  \int \d^d x \alpha   A_\mu^a\frac{\delta S_{YM}}{\delta A_\mu^a}  +  \int \d^d x  \alpha  \left\{  - \p_\mu b^a A_\mu^a  + g f_{akb} A_\mu^k c^b \p_\mu \overline c^a  \right\} \nonumber\\
&& + \int \d^d x \alpha \Bigl[ - g f_{akb} \p_\mu \overline \varphi^a_i A^k_\mu \varphi^b +  g f_{akb} \p_\mu \overline \omega^a_i A^k_\mu \omega^b - g^2 f_{abm} f_{bkd} \p_\mu \overline \omega^a \varphi^m A^k_\mu c^d\Bigr] \nonumber\\
&&+ \int \d^d x \alpha \Bigl[ - g f_{akb} M^a_i A^k_\mu \varphi^b_i  + g f_{akb} U^a_i A^k_\mu \omega^b_i  - g f_{akb} N^a_i A^k_\mu \overline \omega^b_i  - g f_{akb} V^a_i A^k_\mu \overline \varphi^b_i\Bigr] \nonumber\\
&& - \int \d^d x \alpha \left[ g^2 f_{abc} f_{bkd} U^a_i \varphi^c A^k c^d + g^2 f_{abc} f_{bkd} V^a \overline \omega^c A^k c^d \right]\,.
\end{eqnarray}
Notice that we have neglected again higher order terms in the
sources $\sim( \alpha \eta \ldots)$, $\sim ( \alpha \lambda
\ldots)$ and $\sim ( \alpha q \ldots)$ as the argument
\eqref{argument} is still valid. The corresponding counterterm
$\Sigma^{\prime c}$ reads:
\begin{eqnarray}
\Sigma^{\prime c}&=& a_{0}S_{YM} +  b_0 \widehat S_{YM} +
a_{1}\int \d^dx\Biggl(  A_{\mu}^{a}\frac{ \delta S_{YM}}{\delta A_{\mu }^{a}}+ A_{\mu}^{a}\frac{\delta \widehat{S}_{YM}}{\delta A_{\mu }^{a}}  + \p_\mu \overline{c}^a \p_\mu c^a + K_{\mu }^{a}\partial _{\mu }c^{a}  + M_\mu^{a i} \p_\mu \varphi_\mu^{ai} -  U_\mu^{a i} \p_\mu \omega_\mu^{ai} \nonumber\\
&&N_\mu^{a i} \p_\mu \overline{\omega}_\mu^{ai} +  V_\mu^{a i} \p_\mu \overline{\varphi}_\mu^{ai}  +  \p_\mu \overline{\varphi}^{a i} \p_\mu \varphi_\mu^{ai} +  \p_\mu \omega^{a i} \p_\mu \overline{\omega}_\mu^{ai} + V_\mu^{a i} M_\mu^{a i} - U_\mu^{a i}N_\mu^{a i} - g f_{abc} U_\mu^{ia} \varphi^{bi} \p_\mu c^c \nonumber\\
&&- g f_{abc} V_\mu^{ia} \overline{\omega}^{bi} \p_\mu c^c - g f_{abc} \p_{\mu} \overline{\omega}^a \varphi^{bi}  \p_\mu c^c \Biggr)  \nonumber\\
&&+ b_{1}\int \d^dx q \Biggl(  A_{\mu}^{a}\frac{ \delta S_{YM}}{\delta A_{\mu }^{a}}  + \p_\mu \overline{c}^a \p_\mu c^a  + M_\mu^{a i} \p_\mu \varphi_\mu^{ai} -  U_\mu^{a i} \p_\mu \omega_\mu^{ai} +  N_\mu^{a i} \p_\mu \overline{\omega}_\mu^{ai} +  V_\mu^{a i} \p_\mu \overline{\varphi}_\mu^{ai} \nonumber\\
&&+  \p_\mu \overline{\varphi}^{a i} \p_\mu \varphi_\mu^{ai} +  \p_\mu \omega^{a i} \p_\mu \overline{\omega}_\mu^{ai} + V_\mu^{a i} M_\mu^{a i} - U_\mu^{a i}N_\mu^{a i} - g f_{abc} U_\mu^{ia} \varphi^{bi} \p_\mu c^c - g f_{abc} V_\mu^{ia} \overline{\omega}^{bi} \p_\mu c^c - g f_{abc} \p_{\mu} \overline{\omega}^a \varphi^{bi}  \p_\mu c^c \Biggr) \nonumber\\
&&+ a_{1}\int \d^dx \eta \Biggl(  \p_\mu \overline{c}^a \p_\mu c^a  + M_\mu^{a i} \p_\mu \varphi_\mu^{ai} -  U_\mu^{a i} \p_\mu \omega_\mu^{ai} +  N_\mu^{a i} \p_\mu \overline{\omega}_\mu^{ai} +  V_\mu^{a i} \p_\mu \overline{\varphi}_\mu^{ai} \nonumber\\
&&+  \p_\mu \overline{\varphi}^{a i} \p_\mu \varphi_\mu^{ai} +  \p_\mu \omega^{a i} \p_\mu \overline{\omega}_\mu^{ai} + V_\mu^{a i} M_\mu^{a i} - U_\mu^{a i}N_\mu^{a i} - g f_{abc} U_\mu^{ia} \varphi^{bi} \p_\mu c^c - g f_{abc} V_\mu^{ia} \overline{\omega}^{bi} \p_\mu c^c - g f_{abc} \p_{\mu} \overline{\omega}^a \varphi^{bi}  \p_\mu c^c \Biggr) \nonumber\\
&& +  a_{1}\int \d^dx \lambda \Biggl( U_\mu^{a i} \p_\mu \varphi^{a i} +  V_\mu^{a i} \p_\mu \overline \omega^{a i} + \p_\mu \overline \omega^{ai} \p_\mu \varphi^{ai} + U_\mu^{a i} V_\mu^{a i}  \Biggr) - a_{1}\int \d^dx  \Biggl( X^i \p_\mu \overline \omega^{ai} \p_\mu c^a  \Biggr) \nonumber\\
&&+  a_{0}\int \d^dx\Biggl( \alpha A_{\mu}^{a}\frac{\delta S_{YM}}{\delta A_{\mu }^{a}} \Biggr)\nonumber\\
&& + a_1 \int \d^d x \alpha \left( 2 A_\mu^a \p_\mu \p_\nu A_\nu^a - 2 A_\mu^a \p^2 A_\mu^a +
9 g f_{abc} A_\mu^a A_\nu^b \p_\mu A_\nu^c + 4 g^2 f_{abc} f_{cde} A_\mu^a A_\nu^b A_\mu^d A_\nu^e\right)\,,
\end{eqnarray}
once more dropping higher order terms in the sources.\\
\\
The final step in the renormalization procedure is to reabsorb the counterterm $\Sigma^{\prime c}$ into the original action $\Sigma'$,
\begin{equation}
\Sigma (g,\omega ,\phi ,\Phi )+h \Sigma^c = \Sigma (g_{0},\omega_{0},\phi _{0},\Phi_{0})+O(h^{2})\,,  \label{stab}
\end{equation}
We set $\phi =(A_{\mu }^{a}$, $c^{a}$, $\overline{c}^{a}$, $b^{a}$, $\varphi_{i}^{a}$, $\omega_{i}^{a}$, $\overline{\varphi}_{i}^{a}$, $\overline{\omega}_{i}^{a})$ and $\Phi=(K^{a\mu }$, $L^{a}$, $M_{\mu }^{ai}$, $N_{\mu }^{ai}$, $V_{\mu}^{ai}$, $U_{\mu }^{ai}$, $\lambda$) and we define
\begin{align}
g_{0}&=Z_{g}g\,, &  \phi _{0} &=Z_{\phi}^{1/2}\phi \,, & \Phi _{0} &=Z_{\Phi }\Phi \,,
\end{align}
while for the other sources we propose the following mixing matrix
\begin{equation}\label{zmatrixbis}
 \left(
  \begin{array}{c}
    q_0 \\
    \eta_0 \\
    J_0
  \end{array}
\right)=\left(
          \begin{array}{ccc}
            Z_{q      q} & Z_{q      \eta}  & Z_{q      J} \\
            Z_{\eta q} & Z_{\eta \eta}  & Z_{\eta J} \\
            Z_{J      q} & Z_{J      \eta}  & Z_{J      J}
          \end{array}
        \right)
\left(
  \begin{array}{c}
    q \\
    \eta \\
    J
  \end{array}
\right)\,.
\end{equation}
If we try to absorb the counterterm into the original action, we easily find,
\begin{eqnarray}\label{Z1}
Z_{g} &=&1-h \frac{a_0}{2}\,,  \nonumber \\
Z_{A}^{1/2} &=&1+h \left( \frac{a_0}{2}+a_{1}\right) \,,
\end{eqnarray}
and
\begin{eqnarray}\label{Z2}
Z_{\overline{c}}^{1/2} &=& Z_{c}^{1/2} = Z_A^{-1/4} Z_g^{-1/2} = 1-h \frac{a_{1}}{2}\,, \nonumber \\
Z_{b}&=&Z_{A}^{-1}\,, \nonumber\\
Z_{K }&=&Z_{c}^{1/2}\,,  \nonumber\\
Z_{L} &=&Z_{A}^{1/2}\,,
\end{eqnarray}
The results \eqref{Z1} are already known from the
renormalization of the original Yang-Mills action in the Landau
gauge. Further, we also obtain
\begin{eqnarray}\label{Z3}
Z_{\varphi}^{1/2} &=& Z_{\overline \varphi}^{1/2} = Z_g^{-1/2} Z_A^{-1/4} = 1 - h \frac{a_1}{2}\,, \nonumber\\
Z_\omega^{1/2} &=& Z_A^{-1/2} \,,\nonumber\\
Z_{\overline \omega}^{1/2} &=& Z_g^{-1} \,,\nonumber\\
Z_M &=& 1- \frac{a_1}{2} = Z_g^{-1/2} Z_A^{-1/4}\,, \nonumber\\
Z_N &=& Z_A^{-1/2} \,, \nonumber\\
Z_U &=& 1 + h \frac{a_0}{2} = Z_g^{-1} \,, \nonumber\\
Z_V &=& 1- h \frac{a_1}{2} = Z_g^{-1/2}Z_A^{-1/4} \,,
\end{eqnarray}
which are known from the original Gribov-Zwanziger
action, see \cite{Zwanziger:1992qr}. In addition, we also find the
following mixing matrix
\begin{eqnarray}\label{mixingmatrix}
\left(
          \begin{array}{ccc}
            Z_{q      q} & Z_{q      \eta}  & Z_{q      J} \\
            Z_{\eta q} & Z_{\eta \eta}  & Z_{\eta J} \\
            Z_{J      q} & Z_{J      \eta}  & Z_{J      J}
          \end{array}
        \right) &=& \left(
          \begin{array}{ccc}
            1 + h (b_0 - a_0) & 0  & 0 \\
            h b_1 & 1  & 0 \\
            h b_1 & 0  & 1
          \end{array}
        \right)\,,
\end{eqnarray}
while for the $Z$-factor of $\lambda$ we have
\begin{eqnarray}
Z_{\lambda} &=& Z_{c}^{-1/2} Z_A^{-1/2} = Z_g^{1/2} Z_A^{-1/4} \,.
\end{eqnarray}
Also this part was already known, see \cite{Dudal:2008tg}. So far, we have proven that the two limit cases are at least correct. Finally, we find the new results
\begin{eqnarray}
Z_{Y} &=& Z_g Z_A^{-1/2} \,, \nonumber\\
Z_{X} &=& Z_g^{1/2} Z_A^{-1/4}\,.
\end{eqnarray}
In summary, the action $\Sigma'$ is renormalizable. Moreover, we have only 4 arbitrary parameters, $a_0$, $a_1$, $b_0$, $b_1$, which is the same number as in the limit case $\{\varphi, \overline \varphi, \omega, \overline \omega, U, V, N, M \} \to 0$, i.e.~the Yang-Mills case with the introduction of the glueball operator $\sim F^2_{\mu\nu}$ \cite{Dudal:2008tg}. This is already a remarkable fact.

\subsection{Inclusion of the glueball operator in the Refined Gribov-Zwanziger action}
In analogy with \cite{Dudal:2008sp,Dudal:2007cw} we shall add the two dimensional mass term $ \sim \left( \overline{\varphi}^a_i \varphi^a_{i} - \overline{\omega}^a_i \omega^a_i \right)$ to the action $\Sigma_\glue$ in equation \eqref{glueballaction},
\begin{eqnarray}
\Sigma_\Rglue &=& \Sigma_\glue + \Sigma_{\overline{\varphi} \varphi} + \Sigma_{\en}\,,
\end{eqnarray}
whereby
\begin{eqnarray}\label{defex}
\Sigma_{\overline{\varphi} \varphi} &=& \int \d^d x \left( s(-J \overline{\omega}^a_i \varphi^a_{i})\right) =\int \d^d x\left( -J\left( \overline{\varphi}^a_i \varphi^a_{i} - \overline{\omega}^a_i \omega^a_i \right) \right) \,, \nonumber\\
\Sigma_{\en} &=&  \int \d^d x \varsigma \Theta J \,,
\end{eqnarray}
with $J$ and $\theta$ new sources, and $\varsigma$ the parameter already defined in equation \eqref{overeenstemming}. In order to agree with the physical action \eqref{RGZ}, we define the following physical limit,
\begin{align}
 \left. \Theta \right|_{\phys} &=  2 \frac{d (N^2 -1)}{\sqrt{2 g^2 N}} \gamma^2\,.
\end{align}
We further define $sJ = 0$ and $s \Theta =0$, hence the BRST invariance is guaranteed.\\
\\
Let us now investigate the renormalizability of action $\Sigma_\Rglue$. We can go through the same
steps as in the previous section. Therefore, we again add the two external pieces, $S_{\ext,1}$ and $S_{\ext,2}$ as
defined in equation \eqref{Sext1} and \eqref{Sext2}, to the action $\Sigma_\Rglue$
\begin{eqnarray}
\Sigma_\R &=& \Sigma_\Rglue + S_{\ext,1} + S_{\ext,2}\,.
\end{eqnarray}
Subsequently, one can easily  check that all Ward
identities \eqref{wardid1} - \eqref{wardander} and
\eqref{wardideinde} remain unchanged up to potential harmless
linear breaking terms. Therefore, the constraints \eqref{cont1} -
\eqref{cont2} and \eqref{cont3} remain valid. Unfortunately, the
extra integrated Ward identity \eqref{broken1} and the integrated
Ward identity \eqref{broken2} are broken due to the introduction
of the mass term. However, the mass term we have added is not a
new interaction as it is only quadratic in the fields. Therefore,
it cannot introduce new divergences to the massless
theory $\Sigma$, and it can only influence its own
renormalization\footnote{We employ massless renormalization
schemes.} as well as  potentially vacuum terms, i.e.
pure source terms. Also, next to Ward identities \eqref{wardid1} -
\eqref{wardander} and \eqref{wardideinde}, we have a
new identity
\begin{eqnarray}
    \frac{\delta \Sigma_\R }{\delta \Theta} &=&  \varsigma  J\,,
\end{eqnarray}
which is translated to the following constraint  at
the level of the counterterm,
\begin{eqnarray}
    \frac{\delta \Sigma^{c}_\R }{\delta \Theta} &=& 0\,.
\end{eqnarray}
 As a consequence, $\Sigma^{c}_\R$ is independent from
the source $\Theta$. Therefore, it follows that the
form of the counterterm $\Sigma^c_\R$ can be written as
\begin{eqnarray}
    \Sigma^{c}_\R &=& \Sigma^c + \Sigma^c_J\,,
\end{eqnarray}
whereby $\Sigma^c$ is the counterterm \eqref{countertermfinal} of $\Sigma$ and $\Sigma^c_J$ is depending on $J$. One can now
easily check that $\Sigma^c_J = \kappa J^2$, with $\kappa$ a new parameter as this is the only possible combination
with the source $J$, which does not break the constraints \eqref{ST} - \eqref{cont2} and \eqref{cont3}.\\
 \\
$\kappa$ is in fact a redundant parameter, as no divergences in $J^2$ will occur, as explained in \cite{Dudal:2008sp}. Therefore, the counter\-term $\Sigma^c_\R$ is actually equal to $\Sigma^c$. Defining
\begin{align}
   J_{0} &= Z_{J} J \,,
\end{align}
we find
\begin{align}\label{ZJ}
Z_J &= Z_{\varphi}^{-1} = Z_g Z_A^{1/2}\,,
\end{align}
and we have proven the renormalizability of the action $\Sigma^{c \prime}$.

\section{The operator mixing matrix to all orders}
\subsection{Preliminaries}
Let us return to the mixing matrix of the sources $q$, $\eta$ and $J$ and pass to the the corresponding operators. We have found that
\begin{equation}\label{zmatrixbis}
 \left(
  \begin{array}{c}
    q_0 \\
    \eta_0 \\
    J_0
  \end{array}
\right)=\left(
          \begin{array}{ccc}
            Z_{q      q} & 0  & 0 \\
            Z_{J q} & 1  & 0 \\
            Z_{J      q} & 0  & 1
          \end{array}
        \right)
\left(
  \begin{array}{c}
    q \\
    \eta \\
    J
  \end{array}
\right)\,.
\end{equation}
We shall further need the inverse of this matrix,
\begin{equation}\label{zmatrixbis}
 \left(
  \begin{array}{c}
    q \\
    \eta \\
    J
  \end{array}
\right)=\left(
          \begin{array}{ccc}
            \frac{1}{Z_{q    q}} & 0 & 0 \\
            -\frac{Z_{Jq}}{ Z_{q    q}} & 1 & 0 \\
            -\frac{Z_{Jq}}{ Z_{q    q}} & 0  & 1
          \end{array}
        \right)
\left(
  \begin{array}{c}
    q_0 \\
    \eta_0 \\
    J_0
  \end{array}
\right)\,.
\end{equation}
We can write the final action $\Sigma'$ from equation
\eqref{eindactie} in a more condensed form as
\begin{eqnarray}\label{eindactieproper}
 \Sigma' &=& \Sigma_\GZ + S_{\ext,1} + S_{\ext,2} + \int \d^d x  \left( q \mathcal F + \eta \mathcal E + \alpha \mathcal H \right)  + \int \d^d x \lambda \mathcal N\,,
\end{eqnarray}
whereby we have defined the operators
\begin{eqnarray}
\mathcal F &=& \frac{1}{4} F^a_{\mu\nu} F^a_{\mu\nu}\,,  \nonumber\\
\mathcal E &=& s \mathcal N\,, \nonumber\\
\mathcal H &=& A_\mu^a\frac{ S_\GZ}{A_\mu^a}\,,
\end{eqnarray}
with
\begin{eqnarray}
\mathcal N &=& \left[  \p_\mu \overline c^a A_\mu^a + \p \overline \omega \p \varphi + g f_{akb} \p \overline \omega^a A^k \varphi^b + U^a D^{ab} \varphi^b + V^a D^{ab}\overline \omega^b + UV \right]\,.
\end{eqnarray}
It is then an easy task to construct the corresponding mixing matrix for the operators themselves. We recall that insertions of an operator can be obtained by taking derivatives of the generating functional
$Z^c(q,\eta,J)$ w.r.t. to the appropriate source. For example,
\begin{eqnarray}\label{ren1}
    \mathcal{F}_0 &\sim& \frac{\delta Z^c(q, \eta, J)}{\delta q_0} = \frac{\delta q}{\delta q_0}\frac{\delta Z^c(q, \eta, J) }{\delta q}+\frac{\delta \eta}{\delta q_0}\frac{\delta Z^c(q, \eta, J)}{\delta \eta } + \frac{\delta J}{\delta q_0}\frac{\delta Z^c(q, \eta, J)}{\delta J }\,,
\end{eqnarray}
and thus
\begin{eqnarray}
 \mathcal{F}_0 &=&\frac{1}{Z_{q    q}}\mathcal{F}-\frac{Z_{Jq}}{ Z_{q    q}} \mathcal{G}  -\frac{Z_{Jq}}{ Z_{q    q}} \mathcal
    H\,,
\end{eqnarray}
and similarly for $\mathcal{E}_0$ and $\mathcal H_0$. Henceforth, we
find
\begin{eqnarray}\label{operatormatrix}
 \left(
  \begin{array}{c}
    \mathcal F_0 \\
    \mathcal E_0 \\
    \mathcal H_0
  \end{array}
\right) &= & \left(
          \begin{array}{ccc}
            Z_{qq}^{-1}& -Z_{Jq}Z_{qq}^{-1}  &-Z_{Jq}Z_{qq}^{-1} \\
            0  &1 & 0   \\
           0 & 0& 1          \end{array}
        \right)
        \left(
\begin{array}{c}
    \mathcal F \\
    \mathcal E \\
    \mathcal H
  \end{array}
\right)\,.
\end{eqnarray}
This is a nice result as we recover the expected upper triangular form. In addition, as $\mathcal E$ has a $Z$-factor equal to $1$, we also find that the BRST exact operator $\mathcal E$ does not mix with $\mathcal H$, although this mixing would in principle be allowed. This can be understood as follows. The integrated BRST exact operator $\mathcal E$ is in fact proportional to a sum of four (integrated) equations of motion terms and two other terms,
\begin{eqnarray}\label{countcon}
&&\int \d^4 x \Bigl[ \p_\mu b^a A_\mu^a + \p_\mu \overline c^a D^{ab}_{\mu} c^b + \p \overline \varphi \p \varphi - \p \overline \omega \p \omega + g f_{akb} \p \overline \varphi^a A^k \varphi^b  + g f_{akb} \p \overline \omega^a D^{kd} c^d \varphi^b -  g f_{akb} \p \overline \omega^a A^k \omega^b +M_\mu^{ai} D_\mu^{ab} \varphi_i^b \nonumber\\
 &&+ g U_\mu^{ai} f^{abc}  D_\mu^{ab}c^b \varphi_i^c - U_\mu^{ai}  D_\mu^{ab}\omega_i^b  + N_\mu^{ai} D_\mu^{ab} \overline \omega_i^b -gV_\mu^{ai} f^{abc} D_\mu^{bd}c^d \overline \omega_i^c + V_\mu^{ai}  D_\mu^{ab} \overline \varphi_i^b +M_\mu^{ai} V_\mu^{ai}-U_\mu^{ai} N_\mu^{ai}\Bigr] \nonumber\\
 &&= -\int \d^4 x\left( b^a\frac{\delta \Sigma_\GZ }{\delta b^a}+ \overline c^a \frac{\delta \Sigma_\GZ}{\delta \overline c^a} +  \overline \varphi^a\frac{\delta \Sigma_\GZ }{\delta \overline \varphi^a} +  \overline \omega^a\frac{\delta \Sigma_\GZ }{\delta \overline \omega^a} +  M^{ai}_\mu\frac{\delta \Sigma_\GZ }{\delta M^{ai}_\mu} +  U^{ai}_\mu\frac{\delta \Sigma_\GZ }{\delta U^{ai}_\mu} \right)   \,,
\end{eqnarray}
and therefore, like $\mathcal H$, it does not mix with
the other operators. Notice that we can rewrite the integrated
BRST operator in two other forms:
\begin{eqnarray}\label{countcon2}
\eqref{countcon}&=&  -\int \d^4 x\left( b^a\frac{\delta \Sigma_\GZ }{\delta b^a}+ \overline c^a \frac{\delta \Sigma_GZ}{\delta \overline c^a} +   \varphi^a\frac{\delta \Sigma_\GZ }{\delta \varphi^a} + \omega^a\frac{\delta \Sigma_\GZ }{\delta \omega^a} +  N^{ai}_\mu\frac{\delta \Sigma_\GZ }{\delta N^{ai}_\mu} +  V^{ai}_\mu\frac{\delta \Sigma_\GZ }{\delta V^{ai}_\mu} \right)   \,,
\end{eqnarray}
or
\begin{eqnarray}\label{countcon3}
\eqref{countcon}&=&  -\int \d^4 x\left( b^a\frac{\delta \Sigma_\GZ }{\delta b^a}+ c^a \frac{\delta \Sigma_GZ}{\delta c^a} +   \overline \varphi^a\frac{\delta \Sigma_\GZ }{\delta\overline \varphi^a} + \omega^a\frac{\delta \Sigma_\GZ }{\delta \omega^a} +  M^{ai}_\mu\frac{\delta \Sigma_\GZ }{\delta M^{ai}_\mu} +  N^{ai}_\mu\frac{\delta \Sigma_\GZ }{\delta N^{ai}_\mu} \right)   \,.
\end{eqnarray}

\textbf{Remark}\\
We can also use the refined action $\Sigma_\RGZ$ instead of $\Sigma_\GZ$. We define $\Sigma_\RGZ$ as
\begin{eqnarray}\label{RGZSigma}
 \Sigma_\RGZ &=& \Sigma_\GZ + \Sigma_{\overline \varphi \varphi}+  \Sigma_{\en}\,,
\end{eqnarray}
whereby $\Sigma_{\overline \varphi \varphi}$ and $\Sigma_{\en}$
are defined in equation \eqref{defex}. Replacing $\Sigma_\GZ$ by $\Sigma_\RGZ$ does not alter equation
\eqref{operatormatrix}, but it does slightly modify expression
\eqref{countcon},
\begin{equation}\label{countconR}
\int \d^4 x \mathcal E =  -\int \d^4 x\left( b^a\frac{\delta \Sigma_\RGZ }{\delta b^a}+ \overline c^a \frac{\delta \Sigma_\RGZ}{\delta \overline c^a} +  \overline \varphi^a\frac{\delta \Sigma_\RGZ }{\delta \overline \varphi^a} +  \overline \omega^a\frac{\delta \Sigma_\RGZ }{\delta \overline \omega^a} +  M^{ai}_\mu\frac{\delta \Sigma_\RGZ }{\delta M^{ai}_\mu} +  U^{ai}_\mu\frac{\delta \Sigma_\RGZ }{\delta U^{ai}_\mu} - J \frac{\delta \Sigma_\RGZ }{\delta J}  + \Theta \frac{\delta \Sigma_\RGZ }{\delta \Theta }\right)   \,,
\end{equation}
and analogously for expression \eqref{countcon2} and \eqref{countcon3}.

\subsection{The physical limit}
In the next subsection, we shall work in the physical limit as our final intention is to examine $n$-point functions with the (Refined) Gribov-Zwanziger action itself. In the physical limit, $\mathcal E$ becomes:
\begin{multline}\label{mult}
\left.\mathcal E\right|_\phys ~=~  \p_\mu b^a  A_\mu^a + \p_\mu  \overline c^a D_\mu^{ab} c^b +  \p_\mu \overline \varphi_i^a  D_\mu^{ab} \varphi^b_i  - \p_\mu \overline \omega_i^a  D_\mu^{ab} \omega_i^b  + g f^{abc} \p_\mu \overline \omega_i^a    D_\mu^{bd} c^d  \varphi_i^c   + \gamma ^{2} g  f^{abc}A_\mu^a \varphi_\mu^{bc} +  \gamma^2 g f^{abc} A_\mu^a \overline \varphi_\mu^{bc}\\ + d \left(N^{2}-1\right)  \gamma^4 \,.
\end{multline}
From this point, we can omit the constant term $d \left(N^{2}-1\right) \gamma^4$ as it shall not play a role in the calculation of the glueball correlator. Later, we shall determine the renormalization group invariant $\mathcal R(x)$ which contains $F^2_{\mu\nu}(x)$. As $\mathcal E$ mixes with $F^2_{\mu\nu}(x)$, this renormalization group invariant shall also contain this constant term. However, a constant term can never contribute to the final glueball correlator $\braket{\mathcal R(x) \mathcal R (y)}$ as it can never help to produce connected diagrams between the two space time points $x$ and $y$. Therefore, we shall simplify the calculations by omitting this term already from this point.\\
\\
In the physical limit $\mathcal H$ is given by
\begin{eqnarray}\label{H}
\left. \mathcal H\right|_\phys &=& A_\mu^a \frac{\delta S_\GZ}{ \delta A_\mu^a} \,,
\end{eqnarray}
whereby $S_\GZ$ is the physical Gribov-Zwanziger action \eqref{SGZphys}. Naturally, the mixing matrix \eqref{operatormatrix} stays valid.

\subsection{The mixing matrix to all orders}
It this section, we shall determine the mixing matrix \eqref{operatormatrix} to all orders. This proof is very elegant as it does not require to calculate any loop diagrams, and it is purely based on algebraic manipulations. We shall extend the proof given in \cite{Dudal:2008tg}, which is based on \cite{Brown:1979pq}. Moreover, as a byproduct, the proof shall also reveal some identities between the anomalous dimensions of the different fields, which can serve as a check on relations as in \eqref{Z2} and \eqref{Z3}. We shall directly work with the physical action $S_\GZ$. In the end, we shall also look at the Refined Gribov-Zwanziger action, $S_\RGZ$. \\
\\
We start our analysis with the following generic $n$-points function
\begin{eqnarray}\label{definitie}
\mathcal G^n (x_1, \ldots, x_n ) &=& \Braket{\phi_i (x_1)\ldots \phi_j (x_n)  }~=~  \int [\d \phi] \phi_i (x_1)\ldots \phi_j (x_n) \e^{- S_\GZ} \,,
\end{eqnarray}
whereby $\phi_i$, $i = 1\ldots 8$ stands for one of the eight
fields $(A_{\mu }^{a}$, $c^{a}$, $\overline{c}^{a}$, $b^{a}$,
$\varphi_{\mu}^{ab}$, $\omega_{\mu}^{ab}$,
$\overline{\varphi}_{\mu}^{ab}$, $\overline{\omega}_{i}^{a})$,
i.e~$\phi_1 = A_\mu$, $\ldots$, $\phi_8 =
\overline{\omega}^{ab}_{\mu}$. We shall immediately omit the
vacuum term $ \gamma^4 (N^2-1) d$ in the action $S_\GZ$, as it is relevant only for the calculation of the vacuum
energy and not for the calculation of $n$-points functions. The
total number of fields is given by $n$,
\begin{eqnarray}
n &=& \sum_i^8 n_i\,,
\end{eqnarray}
with $n_i$ the number of fields $\phi_i$ present in the $n$-points function \eqref{definitie}. We are therefore considering the path integral for a random combination of fields. Subsequently, from the definition \eqref{definitie}, we can immediately write down the connection between the renormalized Green function and the bare Green function, which is, in a very condensed notation,
\begin{eqnarray}\label{connectie}
\mathcal G^{n} = \prod_{i=1}^8 Z_{\phi_i}^{-n_i/2}  \mathcal G_0^{n}\,.
\end{eqnarray}
From the previous equation, we shall be able to fix all the matrix elements of expression \eqref{operatormatrix}, based on the knowledge that $\frac{\d \mathcal G^{n} }{\d g^2}$ must be finite in a renormalized theory.\\
\\
We shall therefore calculate this quantity. The first step is to apply the chain rule:
\begin{eqnarray}\label{steponeone}
\frac{\d \mathcal G^{n} }{\d g^2} &=& \sum_{j=1}^8 \left( \frac{\p  Z_{\phi_j}^{-n_j/2} }{ \p g^2} \prod_{i \not= j} Z_{\phi_i}^{-n_i/2} \right)     \mathcal G_0^{n} + \prod_{i=1}^8 Z_{\phi_i}^{-n_i/2} \left[ \frac{\p g_0^2 }{ \p g^2} \frac{\p  }{\p g_0^2} +  \frac{\p \gamma_0^2 }{ \p g^2} \frac{\p  }{\p \gamma_0^2} \right] \mathcal G_0^{n}\,.
\end{eqnarray}
Next, we need to calculate the derivatives w.r.t.~$g^2$.
\begin{itemize}
\item Firstly, we need to find $\p g_0^2  / \p g^2$. We employ dimensional regularization, with $d=4-\varepsilon$. If we derive
\begin{eqnarray} \label{glabel}
g_0^2 &=& \mu^\varepsilon Z_g^2 g^2\,,
\end{eqnarray}
w.r.t.~$\mu$ and $g^2$, combine these two equations and employ the following definition of the $\beta$-funtion\footnote{We have immediately extracted the part in $\varepsilon$.}
\begin{eqnarray}
\mu\frac{\p g^2}{\p \mu} &=& - \varepsilon g^2 + \beta(g^2)\,,
\end{eqnarray}
we obtain
\begin{eqnarray}\label{eq1}
\frac{\p g_0^2 }{ \p g^2} &=& \frac{- \varepsilon g_0^2}{ - \varepsilon g^2 + \beta(g^2)}\,.
\end{eqnarray}
\item Secondly, we calculate $ \frac{\p \gamma_0^2 }{ \p g^2}$. We start from
\begin{eqnarray}
\gamma^2_0 &=& Z_{\gamma^2} \gamma^2
\end{eqnarray}
whereby $Z_{\gamma^2} = Z_V =Z_M$ due to the limit
\eqref{physlimit}. Deriving this equation
w.r.t.~$g^2$ yields
\begin{eqnarray}
\frac{\p \gamma^2_0}{\p g^2} &=&  \frac{\p Z_{\gamma^2}}{\p g^2}  \gamma^2 ~=~ \frac{\p \ln Z_{\gamma^2}}{\p g^2}  \gamma^2_0 ~=~  \frac{1}{\mu} \frac{\p \mu}{\p g^2} \mu  \frac{\p \ln Z_{\gamma^2}}{\p \mu}  \gamma^2_0 ~=~\frac{1}{- \varepsilon g^2 + \beta(g^2)} \delta_{\gamma^2} \gamma_0^2\,,
\end{eqnarray}
and we have defined the anomalous dimension of $\gamma^2$ as
\begin{eqnarray}
\delta_{\gamma^2} &=& \mu  \frac{\p \ln Z_{\gamma^2}}{\p \mu} \,.
\end{eqnarray}
\item Finally, we search for $\p  Z_{\phi_j}^{-n_j/2}/ \p g^2$. Applying the chain rule gives
\begin{eqnarray}\label{one}
\frac{\p  Z_{\phi^i}^{-n/2} }{ \p g^2} &=& - \prod_i \frac{ Z_{\phi^i}^{-p_i/2} }{Z_{\phi^i}^{1/2}} \frac{\p Z_{\phi^i}^{1/2}}{ \p g^2} ~=~ - \prod_i Z_{\phi^i}^{-p_i/2} \frac{
\p \ln Z_{\phi^i}^{1/2}}{ \p g^2}\,.
\end{eqnarray}
Next, we derive $ \frac{ \p \ln Z_{\phi^i}^{1/2}}{ \p g^2}$ from the definition of the anomalous dimension,
\begin{eqnarray}\label{two}
\gamma_{\phi^i} &=&\mu \frac{\p \ln Z_{\phi^i}^{1/2}}{\p \mu}  ~=~ \mu \frac{\p g^2 }{\p \mu}  \frac{\p \ln Z_{\phi^i}^{1/2}}{\p g^2} ~=~ \left(  - \varepsilon g^2 + \beta(g^2) \right) \frac{\p \ln Z_{\phi^i}^{1/2}}{\p g^2}\,.
\end{eqnarray}
From expression \eqref{one} and \eqref{two}, it now follows
\begin{eqnarray}\label{eq2}
\frac{\p  Z_{\phi^i}^{-p_i/2} }{ \p g^2} &=& -p_i  Z_{\phi^i}^{-p_i/2} \frac{ \gamma_{\phi^i}}{ - \varepsilon g^2 + \beta(g^2)}\,.
\end{eqnarray}
\end{itemize}
Inserting equation \eqref{eq1} and \eqref{eq2} into expression \eqref{steponeone}, we find:
\begin{eqnarray}\label{stepone}
\frac{\d \mathcal G^{n} }{\d g^2} &=& \frac{ \prod_{i} Z_{\phi_i}^{-n_i/2}}{- \varepsilon g^2 + \beta(g^2)} \left( - \sum_{j=1}^8 n_j  \gamma_{\phi^j} - \varepsilon g_0^2  \frac{\p  }{\p g_0^2} + \delta_{\gamma^2} \gamma^2_0 \frac{\p }{\p \gamma^2_0 }  \right)     \mathcal G_0^{n} \,.
\end{eqnarray}
The right hand side still contains bare and therefore divergent quantities. We would like to rewrite all these quantities in terms of finite quantities so that we can use the finiteness of the left hand side to make observations on the right hand side. Also, we should rewrite in some manner the number $n_i$ as the mixing matrix \eqref{operatormatrix} is obviously independent from these arbitrary numbers.\\
\\
Therefore, as a second step, we shall rewrite the right hand side of \eqref{stepone} in terms of a renormalized quantity. Firstly, we calculate $\frac{\p}{\p g_0^2} \mathcal G_0^{n}$. Using
\begin{equation}
\frac{\p \e^{-S_\GZ}}{\p g_0^2} ~=~-\int \d^4 y \left( - \frac{1}{g_0^2}  \left( \frac{F_0^2(y)}{4}\right)  + \frac{1}{2g_0^2} \left( A_0(y)\frac{\delta S_\GZ}{\delta A_0(y)}- b_0(y) \frac{\delta S_\GZ}{\delta b_0(y) } +  \overline \omega_0(y) \frac{\delta S_\GZ}{\delta \overline \omega_0 (y)} - \omega_0(y) \frac{\delta S_\GZ}{\delta \omega_0(y) } \right)  \right) \e^{-S_\GZ}\,,
\end{equation}
we can write,
\begin{multline}\label{vereenv1}
g_0^2 \frac{\d \mathcal G^{n}_0}{ \d g_0} ~=~  \int \d^4 y \left( \mathcal G^n_0 \left\{ \frac{F_0^2(y)}{4} \right\} - \frac{1}{2}  \mathcal G^n_0 \left\{  A_0(y)\frac{\delta S_\GZ}{\delta A_0(y)} \right\} + \frac{1}{2}  \mathcal G^n_0 \left\{  b_0(y) \frac{\delta S_\GZ}{\delta b_0(y) } \right\} - \frac{1}{2}  \mathcal G^n_0 \left\{  \overline \omega_0(y) \frac{\delta S_\GZ}{\delta \overline \omega_0 (y)} \right\} \right. \\
 \left. +\frac{1}{2}  \mathcal G^n_0 \left\{  \omega_0(y) \frac{\delta S_\GZ}{\delta \omega_0(y)}  \right\}
 \right)\,.
\end{multline}
We have introduced a shorthand notation for an insertion in the $n$-points function, e.g.
\begin{eqnarray}
\mathcal G^{n}_0 \biggl\{ \frac{F_0^2(y)}{4} \biggr\} &=& \Braket{ \frac{F_0^2(y)}{4} \phi^i(x_1)\ldots  \phi^j(z_n) }\,.
\end{eqnarray}
Secondly, we analogously find
\begin{equation} \label{vereenv3}
\gamma_0^2 \frac{\p }{\p \gamma^2_0 } \mathcal G_0^{n} ~=~   \int \d^4 y \left(  \mathcal G^n_0 \left\{  \gamma^2_0 g_0 f^{abc}A_{\mu, 0}^a \varphi_{\mu,0}^{bc} +  \gamma_0^2 g_0 f^{abc} A_{\mu,0}^a \overline \varphi_{\mu,0}^{bc}  \right\} \right)\,.
\end{equation}
Thirdly, we rewrite $n_j \mathcal G^n_0$ by inserting the corresponding counting operator\footnote{It is easily checked that $\int \d^4 y \phi^j_0 \frac{\delta }{\delta \phi^j_0}$ counts the number of $\phi_0^j$ insertions.} into the Green function,
\begin{eqnarray}\label{vereenv2}
n_j \mathcal  G_0^{n}&=&  \int \d^4 y \mathcal G_0^{n}\biggl\{ \phi^j_0(y) \frac{\delta S_\GZ}{\delta \phi^j_0(y) } \biggr\}  \,.
\end{eqnarray}
Inserting \eqref{vereenv1}, \eqref{vereenv3} and \eqref{vereenv2} into our main expression \eqref{stepone} results in
\begin{eqnarray} \label{steptwo}
\frac{\d \mathcal G^{n} }{\d g^2} &=& \frac{1}{- \varepsilon g^2 + \beta(g^2)} \int \d^d y \Biggl[ - \sum_{j=1}^8 \gamma_{\phi^j} \mathcal G^{n}\biggl\{ \phi^j_0(y) \frac{\delta S_\GZ}{\delta \phi^j_0(y) } \biggr\}   - \varepsilon  \mathcal G^n \left\{ \frac{F_0^2(y)}{4} \right\} +  \frac{\varepsilon}{2}  \mathcal G^n \left\{  A_0(y)\frac{\delta S_\GZ}{\delta A_0(y)} \right\} \nonumber\\
 &&- \frac{\varepsilon}{2}  \mathcal G^n \left\{  b_0(y) \frac{\delta S_\GZ}{\delta b_0(y) } \right\} + \frac{\varepsilon}{2}  \mathcal G^n \left\{  \overline \omega_0(y) \frac{\delta S_\GZ}{\delta \overline \omega_0 (y)} \right\}   -\frac{\varepsilon}{2}  \mathcal G^n \left\{  \omega_0(y) \frac{\delta S_\GZ}{\delta \omega_0(y)}  \right\}   \nonumber\\
&&+  \delta_{\gamma^2}  \mathcal G^n \left\{  \gamma^2_0 g_0 f^{abc}A_{\mu, 0}^a \varphi_{\mu,0}^{bc} +  \gamma_0^2 g_0 f^{abc} A_{\mu,0}^a \overline \varphi_{\mu,0}^{bc}  \right\}  \Biggr]\,.
\end{eqnarray}
Notice that we have also absorbed the factor $\prod_{i}
Z_{\phi_i}^{-n_i/2}$ into the Green functions, and therefore we
can replace $\mathcal G_0^n$ again by  $\mathcal G^n$.
Finally, we need to rewrite all the inserted operators in the
$n$-points function $\mathcal G^n$ in terms of their renormalized
counterparts. For this we return to the mixing matrix
\eqref{operatormatrix} and parameterize it as follows
\begin{eqnarray}
 \left(
  \begin{array}{c}
    \mathcal F_0 \\
    \mathcal E_0 \\
    \mathcal H_0
  \end{array}
\right) &= & \left(
          \begin{array}{ccc}
            1 + \frac{a}{\varepsilon}& -\frac{b}{\varepsilon}  &-\frac{b}{\varepsilon} \\
            0  &1 & 0   \\
           0 & 0& 1          \end{array}
        \right)
        \left(
\begin{array}{c}
    \mathcal F \\
    \mathcal E \\
    \mathcal H
  \end{array}
\right)\,.
\end{eqnarray}
Here we have displayed the fact that the entries associated with $a(g^2, \varepsilon)$ and $b(g^2, \varepsilon)$, which represent a
formal power series in $g^2$, must at least have a simple pole in
$\varepsilon$. Therefore, we can rewrite
\begin{eqnarray}
 - \varepsilon  \mathcal F_0(y) &=& \frac{F_0^2(y)}{4} ~=~  \left(  - \varepsilon  - a\right) \mathcal F(y) + b \left. \mathcal E(y) \right|_\phys + b  A(y)\frac{\delta S_\GZ}{\delta A(y)}\,,\nonumber\\
 \left. \mathcal H_0 \right|_\phys &=& A_0(y)\frac{\delta S_\GZ}{\delta A_0(y)} ~=~ A(y)\frac{\delta S_\GZ}{\delta A(y)}\,,
\end{eqnarray}
whereby we recall that we are working in the physical limit and we
have replaced $\left. \mathcal H \right|_\phys$ by the
expression \eqref{H}. Subsequently,
\begin{eqnarray}
 \gamma^2_0 g_0 f^{abc}A_{\mu, 0}^a \varphi_{\mu,0}^{bc} &=&  \gamma^2 g f^{abc}A_{\mu}^a \varphi_{\mu}^{bc}\,, \nonumber\\
 \gamma_0^2 g_0 f^{abc} A_{\mu,0}^a \overline \varphi_{\mu,0}^{bc} &=&   \gamma^2 g f^{abc} A_{\mu}^a \overline \varphi_{\mu}^{bc}\,,
\end{eqnarray}
as one can check with the $Z$-factors in \eqref{Z3}. Finally, all the other operators are equations of motion terms, which appear in expression \eqref{countcon}, \eqref{countcon2} and \eqref{countcon3} and therefore have the same $Z$-factor as the operator $\mathcal E$, i.e.~$Z=1$. Summarizing, expression \eqref{steptwo} becomes:
\begin{eqnarray}\label{steptwobla}
\frac{\d \mathcal G^{n} }{\d g^2} &=& \frac{1}{- \varepsilon g^2 + \beta(g^2)} \int \d^d y \Biggl[ (-\varepsilon - a)\mathcal G^n\left\{ \mathcal F \right\} + \left(\frac{\varepsilon}{2} + b -\gamma_A \right)   \mathcal G^{n} \biggl\{A \frac{\delta S_\GZ}{\delta A } \biggr\}   + \left( -\frac{\varepsilon}{2} -\gamma_b -b \right) \mathcal G^n \left\{  b(y)\frac{\delta S_\GZ}{\delta b(y)} \right\} \nonumber\\
 && \left(- \gamma_{\overline c}  -b \right) \mathcal G^n \left\{  \overline c(y) \frac{\delta S_\GZ}{\delta  \overline c(y)}  \right\} - \gamma_c \mathcal G^n \left\{  c(y) \frac{\delta S_\GZ}{\delta  c(y)}  \right\} +  \left( - \frac{\varepsilon}{2} - \gamma_{\overline \omega} \right)  \mathcal G^n \left\{  \overline \omega(y) \frac{\delta S_\GZ}{\delta \overline \omega (y)} \right\}     \nonumber\\
  &&+\left( \frac{\varepsilon}{2}- \gamma_{ \omega}\right) \mathcal G^n \left\{  \omega(y) \frac{\delta S_\GZ}{\delta \omega(y)}  \right\} -  \gamma_\varphi   \mathcal G^n \left\{  \varphi(y) \frac{\delta S_\GZ}{\delta \varphi (y)}  \right\} -  \gamma_{\overline \varphi}   \mathcal G^n \left\{  \overline \varphi(y) \frac{\delta S_\GZ}{\delta \overline \varphi (y)}  \right\} \nonumber\\
  &&+ b \mathcal G^n \left\{  \p_\mu \overline \varphi_i^a  D_\mu^{ab} \varphi^b_i  - \p_\mu \overline \omega_i^a  D_\mu^{ab} \omega_i^b  + g f^{abc} \p_\mu \overline \omega_i^a    D_\mu^{bd} c^d  \varphi_i^c   + \gamma ^{2} g  f^{abc}A_\mu^a \varphi_\mu^{bc} +  \gamma^2 g f^{abc} A_\mu^a \overline \varphi_\mu^{bc}   \right\}  \nonumber\\
&&+  \delta_{\gamma^2}  \mathcal G^n \left\{  \gamma^2 g f^{abc}A_{\mu}^a \varphi_{\mu}^{bc} +  \gamma^2 g f^{abc} A_{\mu}^a \overline \varphi_{\mu}^{bc}  \right\}  \Biggr]\,.
\end{eqnarray}
where we have immediately taken the full expression of $\left.\mathcal E\right.|_\phys$ in equation \eqref{mult}.\\
\\
From expression \eqref{steptwobla}, we can determine $a(g^2, \varepsilon)$ and $b(g^2, \varepsilon)$. As $\frac{\d \mathcal G^{n} }{\d g^2}$ is a finite expression, we know that the right hand side of equation \eqref{steptwobla} must also be finite. Therefore, as all the Green functions are expressed in terms of finite quantities, we can choose a set of linearly independent terms and demand that their coefficients are finite:
\begin{subequations}
\begin{align}
\mathcal G^n\left\{ \mathcal F \right\} &: \frac{- \varepsilon -a }{-\varepsilon g^2 + \beta(g^2) } \,,  &  \mathcal G^{n} \left\{A \frac{\delta S_\GZ}{\delta A } \right\} &:\frac{ \varepsilon/2 + b - \gamma_A(g^2)  }{-\varepsilon g^2 + \beta(g^2) }\,, \label{a}\\
\mathcal G^n \left\{  b\p_\mu A_\mu \right\} &:\frac{ -\frac{\varepsilon}{2} -\gamma_b -b  }{-\varepsilon g^2 + \beta(g^2) }\,,   & \mathcal G^n \left\{  \overline{c}^{a}\partial _{\mu } D_\mu^{ab} c^b \right\} &: \frac{ - \gamma_{\overline c}  -b - \gamma_c }{-\varepsilon g^2 + \beta(g^2) }\,, \label{b}\\
\mathcal G^n\left\{ \overline \varphi_i^a  \p_\mu D_\mu^{ab}\varphi_i^b \right\} &:  \frac{ -\gamma_\varphi -\gamma_{\overline \varphi} -b  }{-\varepsilon g^2 + \beta(g^2) }\,,  &  \mathcal G^n\left\{ \overline \omega_i^a  \p_\mu D_\mu^{ab} \omega_i^b \right\} &:  \frac{ -\gamma_\omega -\gamma_{\overline \omega} -b  }{-\varepsilon g^2 + \beta(g^2) }\,, \label{c}\\
\mathcal G^n\left\{ -g f^{abc} \p_\nu \overline \omega_i^a  D_\nu^{bd} c^{d} \varphi_i^c \right\} &:  \frac{ -\gamma_c -\gamma_{\overline \omega} - \gamma_\varphi + \frac{\varepsilon}{2} -b} {-\varepsilon g^2 + \beta(g^2) }\,,  &   \label{d}\\
\mathcal G^n\left\{  -\gamma^2 g f^{abc} A_\mu^a \overline \varphi^{bc}  \right\} &:  \frac{  -\gamma_\varphi - \delta_{\gamma^2} - b  }{-\varepsilon g^2 + \beta(g^2) }\,,   & \mathcal G^n\left\{ -\gamma^2 g f^{abc} A_\mu^a \varphi^{bc}  \right\} &:  \frac{   -\gamma_{\overline \varphi} - \delta_{\gamma^2}- b }{-\varepsilon g^2 + \beta(g^2) }\,.  \label{e}
\end{align}
\end{subequations}
We can rewrite the coefficients of $\mathcal G^n\left\{ \mathcal F \right\}$ and $\mathcal G^{n} \left\{A \frac{\delta S_\GZ}{\delta A } \right\}$ in \eqref{a} as
\begin{align}
\frac{- \varepsilon -a }{-\varepsilon g^2 + \beta(g^2) } &= \frac{1}{g^2} \frac{(1  +a/ \varepsilon) }{1 - \beta(g^2) /(\varepsilon g^2) }\,, &  \frac{ \varepsilon/2 + b - \gamma_A(g^2)  }{-\varepsilon g^2 + \beta(g^2) } &= -\frac{1}{2 g^2} \frac{1  + 2( b - \gamma_A(g^2))/\varepsilon }{1 - \beta(g^2) /(\varepsilon g^2) }\,.
\end{align}
Hence, in order to be finite, we must conclude that
\begin{eqnarray}\label{aenb}
a(g^2,\varepsilon) &=& - \frac{\beta(g^2)}{g^2}\,, \nonumber\\
b(g^2,\varepsilon)&=& \gamma_A(g^2) -\frac{1}{2}\frac{\beta(g^2)}{g^2}\,.
\end{eqnarray}
Notice that $a$ and $b$ depends on $g^2$, but not on $\varepsilon$. Therefore, the matrix elements of the first row of the parametrization \eqref{para} only display a simple pole in $\varepsilon$.\\
\\
Moreover, from the other equations we shall obtain relations between the anomalous dimensions of the fields and sources. Let us start with the coefficient of $\mathcal G^n \left\{  b\p_\mu A_\mu \right\}$ in equation \eqref{b}, yielding
\begin{eqnarray}
 \frac{ -\varepsilon/2 - b -  \gamma_b(g^2)  }{-\varepsilon g^2 + \beta(g^2) } &=& \frac{1}{2 g^2} \frac{ 1 + 2(b +  \gamma_b(g^2))/\varepsilon  }{1  - \beta(g^2)/(\varepsilon g^2) }\,,
\end{eqnarray}
which means that
\begin{eqnarray}
b(g^2,\varepsilon)&=&  -  \gamma_b(g^2)  -  \frac{1}{2}  \frac{\beta(g^2)}{g^2}\,.
\end{eqnarray}
Inserting the value of $b(g^2,\varepsilon)$ from expression \eqref{aenb} gives the following relation
\begin{eqnarray}\label{rel}
\gamma_A + \gamma_b &=& 0\,.
\end{eqnarray}
This relation is a translation of the relation $Z_A^{1/2}Z_b^{1/2} = 1$ found in equation \eqref{Z2}. Indeed, deriving both sides w.r.t.~$\mu$ gives
\begin{equation}
 \frac{1}{Z_A^{1/2}Z_b^{1/2}  }\mu \frac{\p}{\p \mu}\left( Z_A^{1/2}Z_b^{1/2} \right) ~= ~ \gamma_A + \gamma_b  ~=~ 0\,.
\end{equation}
Analogously, for the coefficient of $\mathcal G^n \left\{  \overline{c}^{a}\partial _{\mu } D_\mu^{ab} c^b \right\}$, we find
\begin{eqnarray}\label{waardeb}
b(g^2,\varepsilon)&=& -\gamma_c - \gamma_{\overline c}\,,
\end{eqnarray}
yielding
\begin{eqnarray}
\gamma_A + \gamma_c + \gamma_{\overline c} &=& \frac{ \beta }{ 2 g^2}\,,
\end{eqnarray}
which is a translation of $Z_c^{1/2} Z_{\overline c}^{1/2} Z^{1/2}_A Z_g = 1$ as $\mu \frac{\d Z_g}{\d \mu} = - \frac{ \beta }{ 2 g^2}$. Next, the coefficients of \eqref{c} and \eqref{d} lead to
\begin{align}
\gamma_\varphi + \gamma_{\overline \varphi} + \gamma_A & = \frac{\beta}{2 g^2}\,, &  \gamma_\omega + \gamma_{\overline \omega} + \gamma_A & = \frac{\beta}{2 g^2}\,, & \gamma_c + \gamma_{\overline \omega} + \gamma_\varphi + \gamma_A &= \frac{\beta}{g^2}\,,
\end{align}
stemming from
\begin{align}
Z^{1/2}_\varphi Z^{1/2}_{\overline \varphi}Z_A^{1/2} Z_g & = 1\,, &  Z^{1/2}_\omega Z^{1/2}_{\overline \omega} Z^{1/2}_A Z_g & =1 \,, &  Z^{1/2}_c Z^{1/2}_{ \overline \omega} Z_\varphi^{1/2} Z_A^{1/2} Z_g & = 1 \,.
\end{align}
These relations originate from the relations derived in \eqref{Z2} and \eqref{Z3}. Finally, the coefficients in equation \eqref{e} are finite if
\begin{equation}
  -\gamma_{\overline \varphi} - \delta_{\gamma^2} ~=~-\gamma_{ \varphi} - \delta_{\gamma^2} ~=~ b ~= ~\gamma_A(g^2) -\frac{1}{2}\frac{\beta(g^2)}{g^2}\,,
\end{equation}
or equivalently
\begin{align}
Z^{1/2}_{\overline \varphi}Z_A^{1/2} Z_g Z_{\gamma^2} & =1\,, &  Z^{1/2}_{\varphi}Z_A^{1/2} Z_g Z_{\gamma^2} & =1\,,
\end{align}
which is also fulfilled as $Z_{\gamma^2} = Z_V = Z_g^{-1/2} Z_A^{-1/4}$. \\
\\
In summary, we have determined to all orders the mixing matrix \eqref{operatormatrix}. For notational simplicity, we take the value \eqref{waardeb} for $b$ and we use the equality $\gamma_c = \gamma_{\overline c}$:
\begin{eqnarray}\label{para}
Z &= & \left(
          \begin{array}{ccc}
            1 -  \frac{\beta(g^2)}{\varepsilon g^2}& \frac{2\gamma_c}{\varepsilon }  & \frac{2\gamma_c}{\varepsilon }  \\
            0  &1 & 0   \\
           0 & 0& 1          \end{array}
        \right)\,.
\end{eqnarray}
We have encountered numerous checks which show the consistency of our results.\\
\\
\textbf{Remark}\\
This matrix is also valid for the refined action $S_\RGZ$. One can repeat the proof by replacing $S_\GZ$ with $S_\RGZ$ and by adding the following term in $M^2 = J$ to the game,
\begin{eqnarray}
 S_{\overline{\varphi} \varphi} &=& - M^2 \int \d^d x  \left( \overline \varphi^a_i \varphi^a_i - \overline \omega^a_i \omega^a_i \right)\,,
 \end{eqnarray}
see equation \eqref{overeenstemming}. In the end, expression \eqref{steptwobla} will collect an extra term
\begin{eqnarray}
\frac{\d \mathcal G^{n} }{\d g^2} &=& \frac{1}{- \varepsilon g^2 + \beta(g^2)} \int \d^d y \Biggl[ (-\varepsilon - a)\mathcal G^n\left\{ \mathcal F \right\} + \left(\frac{\varepsilon}{2} + b -\gamma_A \right)   \mathcal G^{n} \biggl\{A \frac{\delta S_\RGZ}{\delta A } \biggr\}   + \left( -\frac{\varepsilon}{2} -\gamma_b -b \right) \mathcal G^n \left\{  b(y)\frac{\delta S_\RGZ}{\delta b(y)} \right\} \nonumber\\
 && \left(- \gamma_{\overline c}  -b \right) \mathcal G^n \left\{  \overline c(y) \frac{\delta S_\GZ}{\delta  \overline c(y)}  \right\} - \gamma_c \mathcal G^n \left\{  c(y) \frac{\delta S_\GZ}{\delta  c(y)}  \right\} +  \left( - \frac{\varepsilon}{2} - \gamma_{\overline \omega} \right)  \mathcal G^n \left\{  \overline \omega(y) \frac{\delta S_\GZ}{\delta \overline \omega (y)} \right\}     \nonumber\\
  &&+\left( \frac{\varepsilon}{2}- \gamma_{ \omega}\right) \mathcal G^n \left\{  \omega(y) \frac{\delta S_\GZ}{\delta \omega(y)}  \right\} -  \gamma_\varphi   \mathcal G^n \left\{  \varphi(y) \frac{\delta S_\GZ}{\delta \varphi (y)}  \right\} -  \gamma_{\overline \varphi}   \mathcal G^n \left\{  \overline \varphi(y) \frac{\delta S_\GZ}{\delta \overline \varphi (y)}  \right\} \nonumber\\
  &&+ b \mathcal G^n \left\{  \p_\mu \overline \varphi_i^a  D_\mu^{ab} \varphi^b_i  - \p_\mu \overline \omega_i^a  D_\mu^{ab} \omega_i^b  + g f^{abc} \p_\mu \overline \omega_i^a    D_\mu^{bd} c^d  \varphi_i^c   + \gamma ^{2} g  f^{abc}A_\mu^a \varphi_\mu^{bc} +  \gamma^2 g f^{abc} A_\mu^a \overline \varphi_\mu^{bc}\right\}  \nonumber\\
&&+  \delta_{\gamma^2}  \mathcal G^n \left\{  \gamma^2 g f^{abc}A_{\mu}^a \varphi_{\mu}^{bc} +  \gamma^2 g f^{abc} A_{\mu}^a \overline \varphi_{\mu}^{bc} \right\} +   \delta_{M^2}  \mathcal G^n \left\{ M^2 (\overline \varphi \varphi - \overline \omega \omega)   \right\} \Biggr]\,,
\end{eqnarray}
where we have introduced the anomalous dimension of $M^2$,
\begin{eqnarray}
\delta_{M^2} &=& \mu  \frac{\p \ln Z_{M^2}}{\p \mu} \,.
\end{eqnarray}
This leads to the following extra coefficients
\begin{align}
 \mathcal G^n \left\{ -M^2  \overline \varphi^a_i \varphi^a_i   \right\}&: \frac{- \gamma_{\overline \varphi} -\gamma_{ \varphi} - \delta_{M^2}  }{-\varepsilon g^2 + \beta(g^2) }\,,  &   \mathcal G^n \left\{ M^2 \overline \omega^a_i \omega^a_i \right\}&: \frac{ - \gamma_{\overline \omega} -\gamma_{ \omega} - \delta_{M^2} }{-\varepsilon g^2 + \beta(g^2) }\,.
\end{align}
so that
\begin{align}
  \gamma_{\overline \varphi}+\gamma_{ \varphi}+ \delta_{M^2} &=0\,, &   \gamma_{\overline \omega} +\gamma_{ \omega} + \delta_{M^2} & =0\,,
\end{align}
or equivalently
\begin{align}
  Z^{1/2}_{\overline \varphi}Z^{1/2}_{ \varphi}Z_{M^2} &=1\,, &   Z^{1/2}_{\overline \omega}Z^{1/2}_{ \omega}Z^{1/2}_{M^2} & =1\,,
\end{align}
which is correct as $Z_J = Z_{M^2} = Z_g Z_A^{1/2}$, see equation \eqref{ZJ}. All the other relations stay valid of course.

\section{The glueball correlator}
\subsection{A renormalization group invariant}
As the final step of our analysis, we shall try to
determine a renormalization group invariant operator which
contains $\mathcal F \equiv \frac{F^2_{\mu\nu}(x)}{4}$. This is
useful as we would want to obtain a renormalization group
invariant estimate for the the glueball mass, i.e.~the pole of the corresponding correlator. This analysis is completely
similar to the one presented in \cite{Dudal:2008tg},
due to the fact that the mixing matrix $Z$ is exactly the same.
However, for the benefit of the reader, let us repeat the
analysis. We define the anomalous dimension matrix $\Gamma$ of the
mixing matrix $Z$ as
\begin{eqnarray}
\mu \frac{\p}{\p \mu} Z &=& Z\, \Gamma\,.
\end{eqnarray}
With the following derivatives,
\begin{eqnarray}
\mu \frac{\p}{\p \mu} \left(1 -\frac{\beta / g^2}{\varepsilon}\right) &=& \frac{1}{\varepsilon} ( \varepsilon g^2 - \beta(g^2)) \frac{\p (\beta/g^2)}{\p g^2}\,, \nonumber\\
 \mu \frac{\p}{\p \mu} \frac{2 \gamma_c}{\varepsilon} &=& \frac{1}\varepsilon{}( -\varepsilon g^2 +\beta(g^2)) \frac{\p 2 \gamma_c}{\p g^2}\,,
\end{eqnarray}
we obtain
\begin{eqnarray}\label{Gamma1}
 \Gamma &= & \left(
          \begin{array}{ccc}
            g^2 \frac{\p (\beta / g^2)}{\p g^2}& - 2g^2 \frac{\p \gamma_c}{\p g^2}  &- 2 g^2\frac{ \p  \gamma_c}{\p g^2} \\
            0  &0 & 0   \\
           0 & 0& 0          \end{array}
        \right)\,.
\end{eqnarray}
Notice that this anomalous dimension matrix is finite, as it should be. This matrix $\Gamma$ is related to the anomalous
dimension of the operators, since
\begin{eqnarray}\label{Gamma2}
X_0 ~=~ Z X
&\Rightarrow& 0 ~=~ \mu \frac{\p Z}{ \p \mu}  X + Z \mu \frac{\p X}{\p \mu} \nonumber\\
&\Rightarrow& \mu \frac{\p X}{\p \mu} ~=~ - \Gamma X \,,
\end{eqnarray}
with \begin{eqnarray}
X &=&  \left(
  \begin{array}{c}
    \mathcal F \\
    \mathcal E \\
    \mathcal H
  \end{array}
\right)
\,,\qquad X_{0} ~=~  \left(
  \begin{array}{c}
    \mathcal F_{0} \\
    \mathcal E_{0} \\
    \mathcal H_{0}
  \end{array}
\right)\,.
\end{eqnarray}
We now have all the ingredients at our disposal to determine a
renormalization group invariant operator. We set
\begin{eqnarray}
\mathcal{R}&=& k \mathcal F + \ell  \mathcal E + m
\mathcal H \,,
\end{eqnarray}
with $k$, $\ell$ and $m$ functions of $g^2$, to be chosen in such a way that
\begin{eqnarray}
\mu\frac{\p}{\p\mu}\mathcal{R}&=&\mu \frac{\p k}{\p \mu}  \mathcal F  - k  g^2 \frac{\p (\beta /
g^2)}{\p g^2} \mathcal F  +2k g^2 \frac{\p \gamma_c}{\p g^2}
\mathcal E + 2k g^2 \frac{\p \gamma_c}{\p g^2} \mathcal H  + \mu
\frac{\p \ell }{\p \mu}  \mathcal E   + \mu \frac{\p m }{\p \mu}
\mathcal H   ~=~ 0\,,
\end{eqnarray}
hence
\begin{eqnarray}\begin{cases}
\mu \frac{\p k}{\p \mu}   - k  g^2 \frac{\p (\beta / g^2)}{\p g^2}  =0\,, \nonumber\\
\mu \frac{\p \ell }{\p \mu} + 2k g^2 \frac{\p \gamma_c}{\p g^2} = 0\,, \nonumber\\
\ell = m\,. \end{cases}
\end{eqnarray}
We therefore choose
\begin{eqnarray}\begin{cases}
k(g^2) = \frac{\beta(g^2)}{g^2}\,, \nonumber\\
\ell(g^2)  = m(g^2) = -2\gamma_c(g^2)\,,\end{cases}
\end{eqnarray}
and we conclude that
\begin{eqnarray}\label{RGE}
\mathcal{R}&=& \frac{\beta(g^2)}{g^2} \mathcal F -2\gamma_c(g^2)  \mathcal E  -2\gamma_c(g^2) \mathcal H
\end{eqnarray}
is a renormalization group invariant scalar operator containing
$F_{\mu\nu}^2$, in the case of the Gribov-Zwanziger action
$\Sigma_\GZ$ as well as in the case of the refined action
$\Sigma_\RGZ$.

\subsection{Irrelevance of the terms proportional to the equations of motion}
As we have found a renormalization group invariant, the final goal
\cite{nele} shall be that of evaluating the glueball
correlator
\begin{equation} \label{mainpoint}
\Braket{\mathcal R(x) \mathcal R(y)}_\phys ~=~ \Braket{ \left( \frac{\beta(g^2)}{g^2}  \mathcal F(x) -2 \gamma_c(g^2)  \mathcal E (x) - 2\gamma_c(g^2)  \mathcal H(x) \right) \left( \frac{\beta(g^2)}{g^2}  \mathcal F(y) -2 \gamma_c(g^2)  \mathcal E(y)  - 2\gamma_c(g^2)  \mathcal H(y)\right)}_\phys \,,
\end{equation}
using the (Refined) Gribov-Zwanziger action. However, this is beyond the scope of the present article as this calculation shall be far from trivial, even at lowest order. \\
\\
As usual the equation of motion terms like $\mathcal H$ will not play a role. Let us demonstrate this with a simple example,
\begin{eqnarray}
\Braket{ \mathcal F(x) \mathcal H(y)}_\phys &=& \Braket{  \mathcal F(x) A^a_\mu(y) \frac{\delta S_{\RGZ} }{ \delta A_\mu^a(y)} }~=~ \int [\d \Phi ] \mathcal F(x)  A_\mu^a (y) \frac{\delta S_{\RGZ} }{ \delta A_\mu^a (y)} \e^{-S_{\RGZ}} ~=~  - \int [\d \Phi ] \mathcal F(x) A_\mu^a (y) \frac{\delta  \e^{-S_{\RGZ}} }{ \delta A_\mu^a (y)}  \nonumber\\
&=& \int [\d \Phi ]   \e^{-S_{\RGZ}}  \frac{\delta  \left( A_\mu^a (y) \mathcal F(x) \right) }{ \delta A_\mu^a (y)}  ~=~ \ldots \delta(x-y) + \delta(0) \Braket{\mathcal F(x)}\,,
\end{eqnarray}
which is zero as $x \not= y$ and $\delta(0) = 0$ in dimensional regularization. Therefore, expression \eqref{mainpoint} reduces to,
\begin{multline}
\Braket{\mathcal R(x) \mathcal R(y)}_\phys ~=~ \left( \frac{\beta(g^2)}{g^2} \right)^2 \Braket{  \mathcal F(x) \mathcal F(y) }+ \left(2\gamma_c(g^2) \right)^2 \Braket{  \mathcal E (x) \mathcal E (y)}_\phys \\ -2 \gamma_c(g^2)  \frac{\beta(g^2)}{g^2}  \left(  \Braket{ \mathcal F(x)  \mathcal E(y)}_\phys  +   \Braket{ \mathcal E(x) \mathcal F(y) }_\phys \right)\,.
\end{multline}

\section{Summary and discussion of the relevance of the soft BRST breaking}
In this paper, we have scrutinized the glueball operator $\mathcal F \equiv \frac{F^2_{\mu\nu}}{4}$ using the
(Refined) Gribov-Zwanziger action $S_\GZ$ ($S_\RGZ$). For this, we have followed the
framework of an earlier work \cite{Dudal:2008tg} where we have investigated this operator for the more
simple case of the usual Yang-Mills gauge theory, quantized in the Landau gauge. However, this framework
is heavily based on the existence of the BRST symmetry while neither $S_\GZ$ nor $S_\RGZ$ are BRST invariant \cite{Dudal:2008sp}. Therefore,
throughout the paper, we have relied on the extended model $\Sigma_\GZ$ and $\Sigma_\RGZ$.
With these ``enlarged'' actions, one can then draw very similar conclusions as in the ordinary Yang-Mills case. The results of interest, i.e. those for the (Refined) Gribov-Zwanziger action, then easily follow from these extended models in the physical limit, in which case certain external sources are assigned a suitable value.\\
\\
Firstly, the classically gauge invariant operator $F^2_{\mu\nu}$ mixes with two other operators, a BRST exact operator,
$\mathcal E = s [  \p_\mu \overline c^a A_\mu^a + \p \overline \omega \p \varphi + g f_{akb} \p \overline \omega^a A^k \varphi^b + U^a D^{ab} \varphi^b + V^a D^{ab}\overline \omega^b + UV ]$, and an operator proportional to the gluon equation of motion, $ \mathcal H = A \frac{\delta \Sigma_{\GZ} }{ \delta A} = A \frac{\delta \Sigma_{\RGZ} }{ \delta A}$. By using the algebraic renormalization procedure, we have determined the form of the mixing matrix $Z$ to all orders,
\begin{eqnarray}
 \left(
  \begin{array}{c}
    \mathcal F_0 \\
    \mathcal E_0 \\
    \mathcal H_0
  \end{array}
\right) &= & \left(
          \begin{array}{ccc}
            Z_{qq}^{-1}& -Z_{Jq}Z_{qq}^{-1}  &-Z_{Jq}Z_{qq}^{-1} \\
            0  &1 & 0   \\
           0 & 0& 1          \end{array}
        \right)
        \left(
\begin{array}{c}
    \mathcal F \\
    \mathcal E \\
    \mathcal H
  \end{array}
\right)\,,
\end{eqnarray}
which has an upper triangular form, as required \cite{Collins:1984xc,Collins:1994ee}.\\
\\
In a second part of the paper, we have completely fixed all the elements of this mixing matrix, by using only algebraic arguments. We have found
\begin{eqnarray}
Z &= & \left(
          \begin{array}{ccc}
            1 -  \frac{\beta(g^2)}{\varepsilon g^2}& \frac{2\gamma_c(g^2)}{\varepsilon }  & \frac{2\gamma_c(g^2)}{\varepsilon }  \\
            0  &1 & 0   \\
           0 & 0& 1          \end{array}
        \right)\,,
\end{eqnarray}
which is completely analogous as in the case of the ordinary Yang-Mills theory \cite{Dudal:2008tg}. This is already a remarkable fact. In addition, we have also encountered numerous checks on our results as we have recovered multiple known relations between the anomalous dimensions of all the fields and sources. \\
\\
In the final part, we have determined a renormalization group invariant including $F^2_{\mu\nu}$, given by
\begin{eqnarray}\label{RGE}
 \mathcal{R} &=& \frac{\beta(g^2)}{g^2} \mathcal F -2\gamma_c(g^2)\mathcal E -2\gamma_c(g^2)  \mathcal H \,,
\end{eqnarray}
which is the main result of this paper. This operator would then be a good point to start the study of the (lightest) scalar glueball from, by
means of the correlator $\braket{ \mathcal R(x) \mathcal R(y)}_\phys$ \cite{nele}.\\
\\
In standard Yang-Mills gauge theories, gauge invariant operators
${\cal F}$ only mix with BRST exact and equation of motion type
terms. While the latter always yield trivial
information at the level of correlators, the BRST exact pieces
drop out due to the BRST invariance of the gauge invariant
operator ${\cal F}$ and of the vacuum. In the Gribov-Zwanziger
approach, the situation gets more complicated due to the breaking
of the BRST symmetry\footnote{Recently, it has been shown that it
is nevertheless possible to write down a modified BRST symmetry
generator for the Gribov-Zwanziger action, however at the expense
of allowing for nonlocal transformation behaviour
\cite{Sorella:2009vt,Kondo:2009qz}. This modified BRST generator
reduces to the ordinary one if the Gribov mass $\gamma^2=0$. At
present, it is however unclear if and how such a nonlocal
symmetry, nilpotent \cite{Kondo:2009qz} or not
\cite{Sorella:2009vt}, can be used to define physical operators.}.
In the physical limit, $\mathcal E$ is no longer a BRST invariant
operator. In addition, the BRST symmetry is softly broken.
Therefore, when turning to physical states, $\mathcal E$ will no
longer be irrelevant, and explicitly influence the value of the
correlator. This is not the only observation we can make. ${\cal
R}(x)$ is not the only renormalization group invariant of
dimension 4. Indeed, also the operator $\mathcal {E}(x)$ does not
run with the scale, as we directly infer from equations
\eqref{Gamma1} and \eqref{Gamma2}. We can therefore imagine to
study correlators of linear combinations of the operators ${\cal
F}$ and $\cal E$, where the linear combination is chosen in such a
way that the emerging pole structure would be real. We notice that
this is not a trivial issue in the Gribov-Zwanziger framework
\cite{preprint}, basically due to the fact that the poles of the
gluon propagator itself are already not necessarily real-valued.
When the Gribov parameter $\gamma^2$ is formally set back to zero,
we shall recover the correlators of the usual
kind in Yang-Mills gauge theories, as the BRST symmetry gets restored,
as well as the BRST exactness of the operator $\mathcal{E}$. \\
\\
A research project along the previous lines would thus be very
interesting to pursue. It would also enable us to show
that the soft BRST breaking, deeply related to the presence of the
Gribov horizon, is not necessarily a negative feature of the
theory. Rather, it could be very helpful in the
construction of suitable operators \cite{nele}. We therefore
conclude that the results in this paper have to be seen as a first
step towards the construction of (hopefully) physical correlators
in the GZ theory. As it should have become clear from this paper,
an important tool  has been the possibility of
embedding the (R)GZ theory into the extended model. The nilpotent exact BRST symmetry of the latter model can
be used to identify the renormalizable operators by
using cohomological techniques, which then also give the
renormalizable operators in the physical limit. These latter
operators will contain the classically gauge invariant operators.
At the same time, also renormalizable BRST exact operators can be
found, which reduce to renormalizable operators in the physical
limit, being not necessarily BRST exact. It then
remains to be seen whether suitable linear combinations of these
two types of operators can be found that successfully describe
physical correlators. This will be the topic of future work. As
there are multiple mass scales present in the (Refined)
Gribov-Zwanziger framework, we expect all of them to influence the
pole of the correlators under study \cite{nele}.

\section*{Acknowledgments.}
 We are grateful to L. Baulieu, J.~A.~Gracey and D.~Zwanziger for discussions.
 D.~Dudal and N.~Vandersickel are
 supported by the Research Foundation-Flanders (FWO). The Conselho Nacional de
 Desenvolvimento Cient\'{\i}fico e Tecnol\'{o}gico (CNPq-Brazil), the Faperj,
 Funda{\c{c}}{\~{a}}o de Amparo {\`{a}} Pesquisa do Estado
 do Rio de Janeiro, the SR2-UERJ and the Coordena{\c{c}}{\~{a}}o de
 Aperfei{\c{c}}oamento de Pessoal de N{\'{\i}}vel Superior (CAPES),
 the CLAF, Centro Latino-Americano de F{\'\i}sica, are gratefully acknowledged for financial support.

\end{document}